\documentclass[11pt]{article}
\usepackage{amsfonts}
\usepackage{amssymb}
\usepackage{amsbsy}

\input epsf.sty

\newcommand{\BEQ}{\begin{equation}}     
\newcommand{\BEA}{\begin{eqnarray}}
\newcommand{\EEQ}{\end{equation}}       
\newcommand{\EEA}{\end{eqnarray}}
\newcommand{\del}{\delta}

\newcommand{\loo}{{\mathfrak L}}
\newcommand{\PsiD}{\Psi\!D}

\newcommand{\DPsiD}{D\Psi\!D}
\newcommand{\DPsiDxi}{D\Psi\!D_{\xi}}
\newcommand{\PsiDr}{\Psi\!D_r}
\newcommand{\LDPsiDxi}{{\mathfrak L}_t(D\Psi\!D_{\xi})}
\newcommand{\LPsiDr}{{\mathfrak L}_t((\Psi\!D_r)_{\le 1})}
\newcommand{\LPsiDrtilde}{{\mathfrak L}_t(\widetilde{(\Psi\!D_r)_{\le 1}})}

\newcommand{\g}{{\mathfrak{g}}}

\newcommand{\h}{{\mathfrak{h}}}

\newcommand{\sv}{{\mathfrak{sv}}}

\newcommand{\Tr}{{\mathrm{Tr}}}

\newcommand{\SV}{{\mathrm{SV}}}
\newcommand{\volt}{{\mathfrak{volt}}}

\newcommand{\eop}{$\Box$}
\newcommand{\Vect}{{\mathrm{Vect}}}

\newcommand{\vir}{{\mathfrak{vir}}}

\newcommand{\ad}{{\mathrm{ad}}}

\newcommand{\Diff}{{\mathrm{Diff}}}

\newcommand{\half}{{1\over 2}}
\newcommand{\Del}{\Delta}

\newcommand{\R}{\mathbb{R}}

\newcommand{\C}{\mathbb{C}}
\newcommand{\Z}{\mathbb{Z}}

\newcommand{\II}{{\rm i}}               
\newcommand{\wit}[1]{\widetilde{#1}}    

\catcode`\@=11
\def\numberbysection{\@addtoreset{equation}{section}
        \def\theequation{\thesection.\arabic{equation}}}
\numberbysection

\newtheorem{Theorem}{Theorem}[section]
\newtheorem{Lemma}{Lemma}[section]
\newtheorem{Proposition}[Lemma]{Proposition}

\newtheorem{Definition}[Lemma]{Definition}

\begin{document}

\vskip 1.5 cm
\begin{center}
{\Large \bf A  Hamiltonian action of the Schr\"odinger-Virasoro algebra on a space of periodic time-dependent Schr\"odinger operators
in $(1+1)$-dimensions}
\end{center}

\vskip 2.0 cm
   
\centerline{ {\bf Claude Roger}$^a$ and {\bf J\'er\'emie Unterberger}$^b$}
\vskip 0.5 cm
\centerline {$^a$Institut Camille Jordan ,\footnote{Laboratoire associ\'e au CNRS UMR 5208} 
Ecole Centrale de Lyon, INSA de Lyon, Universit\'e de Lyon, Universit\'e
  Lyon I,} 
\centerline{43 boulevard du 11 novembre 1918, 
F-69622 Villeurbanne Cedex, France}
\vskip 0.5 cm
\centerline {$^b$Institut Elie Cartan,\footnote{Laboratoire 
associ\'e au CNRS UMR 7502} Universit\'e Henri Poincar\'e Nancy I,} 
\centerline{ B.P. 239, 
F -- 54506 Vand{\oe}uvre l\`es Nancy Cedex, France}

\begin{quote}

\renewcommand{\baselinestretch}{1.0}
\footnotesize
{Let ${\cal S}^{lin}:=\{a(t)(-2\II \partial_t-\partial_r^2+V(t,r)\ |\ a\in C^{\infty}(\R/2\pi\Z),
 V\in C^{\infty}(\R/2\pi\Z\times\R)\}$ be the space of Schr\"odinger
operators in $(1+1)$-dimensions with periodic time-dependent potential. The action on ${\cal S}^{lin}$
of a large infinite-dimensional reparametrization
group $SV$ with Lie algebra $\sv$ \cite{RogUnt06,Unt08}, called the Schr\"odinger-Virasoro group and containing the Virasoro group, is proved to be Hamiltonian for a certain Poisson
structure on ${\cal S}^{lin}$. More precisely, the infinitesimal
 action of $\sv$ appears to be part of a  coadjoint action of a Lie algebra of pseudo-differential symbols, $\g$,  of which
$\sv$ is a quotient,
while the Poisson structure is inherited from the corresponding Kirillov-Kostant-Souriau form.
}
\end{quote}

\vspace{4mm}


\newpage


\section{Introduction}


The Schr\"odinger-Virasoro Lie algebra $\sv$ was originally introduced in Henkel \cite{Hen94}
as a natural infinite-dimensional extension of the Schr\"odinger algebra. Recall the latter
is defined as the algebra of projective Lie symmetries of the free Schr\"odinger equation
in (1+1)-dimensions
\BEQ (-2\II {\cal M} \partial_t-\partial_r^2)\psi(t,r)=0. \label{Scheq}\EEQ
These act on equation (\ref{Scheq}) as the following first-order operators
\BEA
L_n=-t^{n+1}\partial_t-\half (n+1)t^n r\partial_r+\frac{\II }{4} {\cal M}
(n+1)nt^{n-1} r^2  - (n+1)\mu t^n \nonumber \\
Y_m=-t^{m+\half}\partial_r+\II {\cal M} (m+\half)t^{m-\half} r  \nonumber\\
M_p= \II {\cal M} t^p \label{repsv0}
\EEA
with $\mu=1/4$ and $n=0,\pm 1$, $m=\pm \half$, $p=0$. The $0$th-order terms in (\ref{repsv0}) correspond on the group
level to the multiplication of the wave function
by a phase. To be explicit, the $6$-dimensional {\it Schr\"odinger group} $\cal S$ acts on $\psi$ by
the following transformations
\BEQ  (L_{-1},L_0,L_1)\ : \quad \psi(t,r)\to \psi'(t',r')=(ct+d)^{-1/2} e^{-\half\II{\cal M} c r^2/(ct+d)} 
\psi(t,r)\label{SchtrL} \EEQ
where $t'=\frac{at+b}{ct+d}$, $r'=\frac{r}{ct+d}$ with $ad-bc=1$; 
\BEQ
(Y_{\pm \half})\ : \quad \psi(t,r)\to \psi(t,r')=e^{-\II{\cal M}  (vt+r_0)(r-v/2)} \psi(t,r) 
\label{SchtrY} \EEQ
where $r'=r-vt-r_0$; 
\BEQ
(M_0)\ :\quad  \psi(t,r)\to e^{\II{\cal M} \gamma } \psi(t,r). \label{SchtrM} \EEQ

The  Schr\"odinger group is isomorphic to a semi-direct  product of $SL(2,\R)$ (corresponding to time-reparametrizations (\ref{SchtrL}))
by the Heisenberg group ${\cal H}_1$ (corresponding to the Galilei transformations
(\ref{SchtrY}), (\ref{SchtrM})). Note that the last transformation (\ref{SchtrM}) (multiplication by a constant phase) is generated by the commutators of the Galilei transformations
(\ref{SchtrY}) - these do not commute because of the added phase terms, which produce a central extension.

The free Schr\"odinger equation comes out naturally when considering many kinds of problems
in out-of-equilibrium statistical physics. Its analogue in equilibrium statistical physics is the
Laplace equation $\Del \psi=0$. In two-dimensional space, the latter equation is invariant by local
conformal transformations which generate (up to a change of variables) the well-known (centerless) Virasoro algebra $\Vect(S^1)$, otherwise
known as the Lie algebra of $C^{\infty}$-vector fields on the torus $S^1:=\{e^{\II\theta},\theta\in[0,2\pi]\}$.
 There is no substitute
for $\Vect(S^1)$ when  time-dependence is included, but the {\it Schr\"odinger-Virasoro Lie algebra}

\BEQ \sv\simeq \langle L_n,Y_m,M_p\ |\ n,p\in\Z, m\in\half+\Z\rangle \EEQ
shares some properties with it. First, the Lie subalgebra span$(L_n,\ n\in\Z)$ is isomorphic to
$\Vect(S^1)$. Actually, $\sv$ is isomorphic to a semi-direct product of  $\Vect(S^1)$ by an
infinite-dimensional rank-two nilpotent Lie algebra.
Second, there exists a family of  natural
 actions of the Schr\"odinger-Virasoro group $SV$ integrating $\sv$  (see \cite{RogUnt06}) on the
  space ${\cal S}^{lin}:=\{a(t)(-2\II {\cal M}\partial_t-
\partial_r^2)+V(t,r)\}$ 
of Schr\"odinger operators  with time-periodic potential, which generalizes the well-known action
$\phi_*: \partial_t^2+u(t)\to \partial_t^2 + (\dot{\phi}(t))^2 (u\circ\phi)(t)+\half S(\phi)(t)$  
(where $S$ stands for  Schwarzian derivative, see below) of the Virasoro group
on
Hill operators. The family of infinitesimal actions of $\sv$ on ${\cal S}^{lin}$, denoted by
$d\tilde{\sigma}_{\mu}$, $\mu\in\R$,
 is introduced in section 1.
 It is essentially obtained by conjugating Schr\"odinger operators with
the above functional transformations  (\ref{repsv0}).

The main result of this paper is the following (see Theorem \ref{th:th}).

{\bf Theorem.}

{\it
There exists a Poisson structure on ${\cal S}^{lin}=\{a(t)(-2\II{\cal M}\partial_t-\partial_r^2)+V(t,r)\}$ for which 
the infinitesimal action $d\tilde{\sigma}_{\mu}$ of $\sv$ is Hamiltonian.
}

The analogue in the case of  Hill operators is well-known (see for instance \cite{GuiRog}). Namely,
the action of the Virasoro group on the space $\cal H$ of Hill operators is equivalent to its affine
coadjoint action  with central charge $c=\half$, with the identification $\partial_t^2+u(t)\to
u(t)dt^2\in \vir^*_{\half}$, where $\vir^*_c$ is the affine hyperplane $\{(X,c)\ |\ X\in\Vect(S^1)^*\}$.
  Hence this action preserves the  canonical KKS
(Kirillov-Kostant-Souriau) structure on $\vir^*_{\half}\simeq{\cal H}$.
As well-known, one may exhibit  a bi-Hamiltonian structure on $\vir^*$ which provides an
integrable system on $\cal H$ associated to the Korteweg-De Vries equation. 

The above identification does not hold true any more in the case of the Schr\"odinger action
of $SV$ on the
space of Schr\"odinger operators, which is {\it not}  equivalent to its coadjoint action (see \cite{RogUnt06}, section
3.2). Hence
the existence of  a Poisson structure for which the  action on Schr\"odinger operators  is Hamiltonian has to
be proved in the first place. It turns out that the action on Schr\"odinger operators is part of
the  coadjoint action of a much larger Lie algebra
 $\g$ on its dual. The  Lie algebra $\g$  is
introduced  in Definition \ref{def:g}. 

The way we went until we came across this Lie algebra $\g$
is a bit tortuous. 

The first idea (see \cite{RogUnt06}, or \cite{HenUnt06} for superized versions of this statement)
 was to see $\sv$ as
a {\it subquotient} of an algebra $\DPsiD$  of {\it extended} pseudodifferential symbols on the line: one easily checks 
that the assignment ${\cal L}_f\to -f(\xi)\partial_{\xi}$, ${\cal Y}_g\to -g(\xi)\partial_{\xi}^{\half}$,
${\cal M}_h\to -\half h(\xi)$ yields a linear application  $\sv\to\DPsiD:=\R[\xi,\xi^{-1}]]\
 [\partial_{\xi}^{\half},\partial_{\xi}^{-\half}]]$ which respects the Lie brackets of both Lie
algebras, up to  unpleasant terms which are pseudodifferential symbols of {\it negative} order. 
Define $\DPsiD_{\le \kappa}$ as the subspace of pseudodifferential symbols with order $\le \kappa$.
Then $\DPsiD_{\le 1}$ is a Lie subalgebra of $\DPsiD$, $\DPsiD_{\le -\half}$ is an ideal, and
the above assignment defines an isomorphism $\sv\simeq \DPsiD_{\le 1}/\DPsiD_{\le -\half}$.

The second idea (sketched in \cite{Unt06}) 
was to use  a non-local transformation $\Theta:\DPsiD\to\PsiD$ ($\PsiD$ being the
usual algebra of pseudo-differential symbols)  which maps $\partial_{\xi}^{\half}$
to $\partial_r$ and $\xi$  to $\half r\partial_r^{-1}$
 (see Definition \ref{def:Theta}). The transformation $\Theta$ is formally an integral
operator, simply associated to the heat
kernel, which maps the first-order differential operator $-2\II{\cal M}\partial_t-\partial_{\xi}$ into
 $-2\II{\cal M}\partial_t-\partial_r^2$. The operator $-2\II{\cal M}\partial_t-\partial_{\xi}$ (which is
simply the  $\partial_{\bar{z}}$-operator in complex coordinates) is now easily seen to be invariant under
an infinite-dimensional Lie algebra which generates (as an associative algebra) an algebra isomorphic
to $\DPsiD$. One has thus defined a natural action of $\DPsiD$ on the space of solutions of the free Schr\"odinger
equation $(-2\II{\cal M}\partial_t-\partial_r^2)\psi=0$. 

The crucial point  now is that (after conjugation with $\Theta$, i.e. coming back to the
usual $(t,r)$-coordinates) the action of $\DPsiD_{\le 1}$ coincides {\it up to pseudodifferential symbols
of negative order} with the above realization (\ref{repsv0})
of the generators $L_n$, $Y_m$, $M_p$ $(n,p\in\Z,m\in\half+\Z)$. In other words, loosely speaking,
the abstract isomorphism $\sv\simeq \DPsiD_{\le 1}/\DPsiD_{\le -\half}$ has received a concrete
interpretation, and one has somehow reduced a problem concerning {\it differential operators in two
variables} $t,r$ into a problem concerning {\it time-dependent pseudodifferential operators in one
space
variable}, which is a priori much simpler.

 Integrable systems associated to Poisson structures on
the loop algebra $\loo_t(\PsiD)$ over $\PsiD$ (with the  usual Kac-Moody cocycle $(X,Y)\to \oint \Tr \dot{X}(t) Y(t)\ dt$,
where $\Tr$ is Adler's trace on $\PsiD$) have been studied by A. G. Reiman and M. A.
Semenov-Tyan-Shanskii \cite{Rei84}.  In our case, computations show that  the $\sv$-action on  Schr\"odinger operators is related to the
coadjoint action of $\LPsiDrtilde$, where $\LPsiDrtilde$ is a central extension of $\LPsiDr$ which is
unrelated to the Kac-Moody cocycle.

Actually, the
 above scheme works out perfectly fine only for the restriction of the $\sv$-action to the nilpotent part of $\sv$. For
reasons explained in sections 3 and 4, the generators of $\Vect(S^1)\hookrightarrow\sv$ play a particular r\^ole. 
So the action $d\tilde{\sigma}_{\mu}$ of $\sv$ is really obtained as part 
of the coadjoint action of an extended Lie algebra $\g:=\Vect(S^1)\ltimes\LPsiDrtilde$, where
$\Vect(S^1)$ acts as the time-dependent outer derivations $f(t)\partial_t, f\in C^{\infty}(S^1)$
on the loop algebra $\LPsiDrtilde$.

It is natural to expect that there should exist some bi-Hamiltonian structure on ${\cal S}^{lin}$
allowing to define some unknown integrable system. We hope to answer this question in the future.

It appears in the
course of the computations that the very closely related 
family of {\em affine} actions $d\sigma_{\mu}$ on ${\cal S}^{aff}:=\{-2\II{\cal M}
\partial_t-\partial_r^2+V(t,r)\}\subset{\cal S}^{lin}$, originally defined in \cite{RogUnt06} (see also Remark in section 1
below), although seemingly
more natural than the linear actions $d\tilde{\sigma}_{\mu}$, is {\em not} Hamiltonian for the same
Poisson structure. 
Note however that the action $d\sigma_{1/4}$ restricted to the 
affine subspace ${\cal S}^{aff}_{\le 2}:=\{-2\II{\cal M}\partial_t
-\partial_r^2+V_2(t)r^2+V_1(t)r+V_0(t)\}$ has been shown in \cite{Unt08} 
to be Hamiltonian for a totally different Poisson structure. The two constructions are unrelated. 

\bigskip

Here is an outline of the article. The definitions and results from \cite{RogUnt06} needed on the Schr\"odinger-Virasoro
algebra and its action on Schr\"odinger operators are briefly recalled in section 1.  Section 2 on pseudo-differential
operators is mainly introductive, except for the definition of the non-local transformation  $\Theta$. The
realization of $\DPsiD_{\le 1}$ as symmetries of the free Schr\"odinger equation is explained in section 3.  Sections
4 and 5 are devoted to the construction of the extended Lie algebra $\LPsiDrtilde$ and its extension
by derivations, $\g$.  The action
 $d\tilde{\sigma}_{\mu}$ of $\sv$ on
Schr\"odinger operators is obtained as part of the coadjoint  action of $\g$ restricted
 to a stable submanifold ${\cal N}\subset\g^*$ defined in section 6, where the main  theorem is stated and
proved. Finally, an explicit rewriting in terms of the underlying Poisson formalism is given in
  section 7.

\underline{Notation:} In the sequel, the derivative with respect to $r$, resp. $t$  will always be denoted
 by a prime $(')$, resp. by a dot, namely,
$V'(t,r):=\partial_r V(t,r)$ and $\dot{V}(t,r):=\partial_t V(t,r)$ (except  the third-order time derivative
$\frac{d^3 V}{dt^3}$, for typographical reasons).


\section{Definition of the action of $\sv$ on Schr\"odinger operators}


We recall in this preliminary section the properties of the Schr\"odinger-Virasoro algebra $\sv$ proved
in \cite{RogUnt06} that will be needed throughout the article.

We shall denote by $\Vect(S^1)$ the Lie algebra of $2\pi$-periodic $C^{\infty}$-vector fields. It is generated by
$(\ell_n;\ n\in\Z)$, $\ell_n:=\II e^{\II n\theta} \partial_{\theta}$, with the following Lie brackets: $[\ell_n,\ell_p]
=(n-p)\ell_{n+p}$. Setting $t=e^{\II\theta}\in S^1$, one has $\ell_n=-t^{n+1}\partial_t$. It may be seen as the Lie
algebra of $\Diff(S^1)$, which is the group of orientation-preserving smooth diffeomorphisms of the torus.

For any $\mu\in\R$, $\Diff(S^1)$ admits a representation on the space of {\em $(-\mu)$-densities}
$${\cal F}_{\mu}:=\{f(\theta)(d\theta)^{-\mu},\quad f\in C^{\infty}(\R/2\pi\Z)\}$$
defined as the natural action by change of variables,
$$\pi_{\mu}(\phi^{-1})f=(\dot{\phi})^{-\mu} f\circ\phi.$$

As well-known, the contragredient representation $({\cal F}_{\mu})^*$ is isomorphic to 
${\cal F}_{-1-\mu}$; a particular case of this is the well-known isomorphism $\Vect(S^1)^*\simeq
{\cal F}_1^*\simeq
{\cal F}_{-2}$, so an element of the restricted dual $\Vect(S^1)^*$ may be represented as 
a tensor density $f(\theta)d\theta^2$, $f\in C^{\infty}(S^1)$.

\begin{Definition}[Schr\"odinger-Virasoro algebra](see \cite{RogUnt06}, Definition 1.2)

{\it
 We denote by $\sv$  the Lie algebra with generators
$
L_n,Y_m,M_n (n\in\Z,m\in\half+\Z)$ and
  following relations (where $n,p\in\Z,m,m'\in\half+\Z$) :
$$
[L_n,L_p]=(n-p)L_{n+p}
$$
$$
[L_n,Y_m]=({n\over 2}-m)Y_{n+m},\quad [L_n,M_p]=-pM_{n+p};
$$
$$
[Y_m,Y_{m'}]=(m-m')M_{m+m'},
$$
$$  [Y_m,M_p]=0,\quad  [M_n,M_p]=0.
$$

If $f$ (resp. $g$, $h$) is a Laurent series, $f=\sum_{n\in \Z} f_n t^{n+1}$, resp. $g=\sum_{n\in\half+\Z} g_n t^{n+\half}$,
$h=\sum_{n\in\Z}  h_n t^n$, then we shall write
\BEQ {\cal L}_f=\sum f_n L_n,\quad {\cal Y}_g=\sum g_n Y_n,\quad {\cal M}_h=\sum h_n M_n.\EEQ
}

\end{Definition}

Let $\g_0=$span($L_n,\ n\in\Z$) and $\h=$span($Y_m,M_p,\ m\in\half+\Z,p\in\Z$). Then $\g_0\simeq\Vect(S^1)$ and
$\h$ are Lie subalgebras of $\sv$, and $\sv\simeq \g_0\ltimes\h$ enjoys a semi-direct product structure. Note also
that $\h$ is rank-two nilpotent.

The Schr\"odinger-Virasoro algebra may be exponentiated into a group $SV=G_0\ltimes H$, where $G_0\simeq
\Diff(S^1)$ and $H$ is a nilpotent Lie group (see \cite{RogUnt06}, Theorem 1.4).

\begin{Definition} (see \cite{RogUnt06}, Definition 1.3)

{\it
  Denote by $d\pi_{\mu}$ the representation
 of $\sv$ as differential operators of order one
on $\R^2$ with coordinates $t,r$ defined by
\BEA
d\pi_{\mu}({\cal L}_f)=-f(t)\partial_t-\half \dot{f}(t) r\partial_r+\frac{1}{4}\II{\cal M}
\ddot{f}(t) r^2  - \mu \dot{f}(t) \nonumber \\
d\pi_{\mu}({\cal Y}_g)=-g(t)\partial_r+\II{\cal M} \dot{g}(t) r  \nonumber\\
d\pi_{\mu}({\cal M}_h)=\II  {\cal M} h(t)  \label{repsv}
\EEA
}

\label{def:pi}

\end{Definition}

Note that $d\pi_{\mu}(L_n)$, $d\pi_{\mu}(Y_m)$, $d\pi_{\mu}(M_p)$ coincide with the formulas
(\ref{repsv0}) given in the Introduction.

The infinitesimal representation $d\pi_{\mu}$ of $\sv$ may be exponentiated into 
a representation $\pi_{\mu}$ of the group $SV$ (see \cite{RogUnt06}, Proposition 1.6). 
 Let us simply write out the exponentiated action \cite{Unt08} :
\BEQ (\pi_{\mu}(\phi;0)f)(t',r')=(\dot{\phi}(t))^{-\mu} e^{\frac{{\cal M}}{4} \II \frac{\ddot{\phi}(t)}{\dot{\phi}(t)}
r^2} f(t,r) \label{eq:pimuphi}  \EEQ
if $\phi\in G_0\simeq\Diff(S^1)$ induces the coordinate change $(t,r)\to (t',r')=(\phi(t),
r\sqrt{\dot{\phi}(t)})$; and
\BEQ (\pi_{\mu}(1;(\alpha,\beta))f)(t',r')=e^{-\II{\cal M}(\dot{\alpha}(t)r-\half \alpha(t)\dot{\alpha}(t)+\beta(t))}
f(t,r) \EEQ
if $(\alpha,\beta)\in C^{\infty}(\R/2\pi\Z)\times C^{\infty}(\R/2\pi\Z)$ induces the coordinate
change $(t,r)\to (t,r')=(t,r-\alpha(t))$.

It appears clearly in eq. (\ref{eq:pimuphi})
 that the parameter $\mu$ is a 'scaling dimension' or the weight of a density.


\bigskip

Let us now introduce the manifold ${\cal S}^{lin}$ of Schr\"odinger operators we want to consider,
and also the affine subspace ${\cal S}^{aff}$.

\begin{Definition}[Schr\"odinger operators] (see \cite{RogUnt06}, Definition 2.1)

{\it
 Let ${\cal S}^{lin}$ be the vector space of second order
operators on $\R^2$ defined by
$$D\in{\cal S}^{lin}\Leftrightarrow D=a(t)(-2\II{\cal M}\partial_t-\partial_r^2)
+V(t,r),
\quad a\in C^{\infty}(\R/2\pi\Z),\ V\in C^{\infty}(\R/2\pi\Z \times \R)$$
and ${\cal S}^{aff}\subset {\cal S}^{lin}$ the affine subspace of
'Schr\"odinger operators' given by the hyperplane $a=1$.

In other words, an element of ${\cal S}^{aff}$ is the sum of the free
Schr\"odinger
operator
$\Del_0:=-2\II{\cal M}\partial_t-\partial_r^2$ and of a time-periodic potential $V$.
}

\end{Definition}

The action of $SV$ on Schr\"odinger operators is essentially the conjugate action of $\pi_{1/4}$
(see following Proposition and Remark) :

\begin{Proposition} (see \cite{RogUnt06}, Proposition 2.5, Proposition 2.6)

{\it

\begin{enumerate}
\item

Let $\tilde{\sigma}_{\mu}: SV\to Hom({\cal S}^{lin},{\cal S}^{lin})$ the representation
of the group of SV on the space of Schr\"odinger operators defined by the left-and-right
action
$$\tilde{\sigma}_{\mu}(g): D\to \pi_{\mu+2}(g) D \pi_{\mu}(g)^{-1},\quad
g\in SV, D\in {\cal S}^{lin}.$$
Then the action of $\tilde{\sigma}_{\mu}$  is given by the following formulas {\em \`a v\'erifier!}:
\BEA
&&\tilde{\sigma}_{\mu}(\phi;0).(a(t)(-2\II{\cal M} \partial_t-\partial_r^2)+V(t,r))=\nonumber\\
&&\qquad\qquad 
\dot{\phi}(t)a(\phi(t))(-2\II{\cal M}  \partial_t-\partial_r^2)+\dot{\phi}^2(t) V(\phi(t),r\sqrt{\dot{\phi}(t)}) +
a\left( 
2\II(\mu-\frac{1}{4}){\cal M} \frac{\ddot{\phi}}{\dot{\phi}} + \half {\cal M}^2
  r^2 S(\phi)(t) \right) \nonumber \\
&& \tilde{\sigma}_{\mu}(1;(\alpha,\beta)).(-2\II{\cal M} \partial_t-\partial_r^2+V(t,r))=\nonumber \\
&& \qquad\qquad -2\II{\cal M} \partial_t-\partial_r^2+V(t,r-\alpha(t))+a\left( -2{\cal M}^2  r\ddot{\alpha}(t)- {\cal M}^2
(2\dot{\beta}(t)-\alpha(t)\ddot{\alpha}(t)) \right)
\EEA

where $S:\phi\to \frac{d^3 \phi/dt^3}{\dot{\phi}}-\frac{3}{2} \left(\frac{\ddot{\phi}}{\dot{\phi}}\right)^2$ is the Schwarzian derivative. 

\item

Let $\Del_0:=-2\II{\cal M}\partial_t-\partial_r^2$ be the free Schr\"odinger operator. 
The infinitesimal action $d\tilde{\sigma}_{\mu}: X\to \frac{d}{dt}\big|_{t=0} \left(\tilde{\sigma}_{\mu}(\exp tX)\right)$ of $\sv$  writes
(recall $V':=\partial_r V$) :

\BEA
&& d\tilde{\sigma}_{\mu}({\cal L}_f)(a(t)\Del_0+V(t,r))=\nonumber\\
&& \qquad -(a\dot{f}+f\dot{a})\Del_0 -f\dot{V}-\half \dot{f}rV'+a\left(
-2\II(\mu-\frac{1}{4})
{\cal M}\ddot{f}-\frac{1}{2}{\cal M}^2 \frac{d^3 f}{dt^3} r^2\right)-2\dot{f}V \nonumber\\
&& d\tilde{\sigma}_{\mu}({\cal Y}_g)(a(t)\Del_0+V(t,r))=-gV'-2{\cal M}^2 a\ddot{g}r \nonumber\\
&& d\tilde{\sigma}_{\mu}({\cal M}_h)(a(t)\Del_0+V(t,r))=-2{\cal M}^2 a\dot{h}
\EEA

\end{enumerate}
}

\label{prop:sigma}

\end{Proposition}

\medskip

{\bf Remark.}

\smallskip

Consider instead the left-and-right action
\BEQ \sigma_{\mu}(g):D\to \pi_{\mu+1}(g)D\pi_{\mu}(g)^{-1},\quad g\in SV,D\in {\cal S}^{lin}.\EEQ
The restriction $\sigma_{\mu}\big|_H$ to the nilpotent subgroup coincides with $\tilde{\sigma}_{\mu}\big|_H$, while
\BEQ d\sigma_{\mu}({\cal L}_f)=-f\dot{a}\Del_0-f\dot{V}-\half \dot{f}rV'+a
\left(
-2\II(\mu-\frac{1}{4})
{\cal M}\ddot{f}-\frac{1}{2}{\cal M}^2 \frac{d^3 f}{dt^3} r^2\right)-\dot{f}V.\EEQ
Hence  $\sigma_{\mu}$ restricts to an affine action on the affine subspace ${\cal S}^{aff}:=\{-2\II{\cal M}\partial_t
-\partial_r^2+V(t,r)\}\subset {\cal S}^{lin}$ corresponding to a constant coefficient $a\equiv 1$.
It appears somehow in the computations that one obtains as a by-product of a certain coadjoint action the family
of linear representations $d\tilde{\sigma}_{\mu}$, and not the affine representations $d\sigma_{\mu}$.

The affine action $d\sigma_{\mu}$ has been studied elsewhere \cite{Unt08} in the
case $\mu=\frac{1}{4}$. (Note that this case is the 'optimal one' as appears in the
formula for $d\sigma_{\mu}({\cal L}_f)$ in Proposition \ref{prop:sigma}: in particular,
only for
$\mu=\frac{1}{4}$ is the {\em free} Schr\"odinger equation preserved by the Schr\"odinger
group, which may be  interpreted by saying that the scaling dimension of the Schr\"odinger
field is $\frac{1}{4}$.)   Once restricted to the stable
submanifold ${\cal S}_{\le 2}^{aff}:=\{-2\II{\cal M}\partial_t-\partial_r^2+V_0(t)+V_1(t)r+V_2(t)r^2\}$ of Schr\"odinger
operators with time-dependent {\it quadratic} potential, it exhibits a rich variety of finite-codimensional orbits,
whose classification is obtained by generalizing classical results due to A. A. Kirillov on orbits of the space of Hill
operators under the Virasoro group. Also, a parametrization of operators by their stabilizers yields a natural
symplectic structure for which the $\sigma_{1/4}$-action is Hamiltonian. These ideas do not carry over to
the whole space ${\cal S}^{lin}$, whose Poisson structure will be obtained below by a totally different method.


\section{Algebras of pseudodifferential symbols}


\begin{Definition}[algebra of formal pseudodifferential symbols]
{\it Let $\PsiD:=\R[z,z^{-1}]] \ [\partial_z,\partial_z^{-1}]]$ be the associative algebra of Laurent series in $z$, $\partial_{z}$
with defining relation $[\partial_z,z]=1$.
}
\end{Definition}

Using 
 the  coordinate $z=e^{\II \theta}$, $\theta\in\R/2\pi\Z$, one may see elements
of $\PsiD$ as formal pseudodifferential operators with periodic coefficients.

The algebra $\PsiD$ comes with a trace, called {\it Adler's trace}, defined in the Fourier coordinate $\theta$ by
\BEQ \Tr \left( \sum_{q=-\infty}^{N} f_q(\theta)\partial_{\theta}^q \right)
=\frac{1}{2\pi} \int_0^{2\pi} f_{-1}(\theta) \ d\theta.\EEQ
Coming back to the coordinate $z$, this is equivalent to setting

\BEQ \Tr(a(z)\partial_z^q)=\del_{q,-1}\ .\ \frac{1}{2\II \pi} \oint a(z)dz\EEQ
where $\frac{1}{2\II\pi}\oint $
 is the Cauchy integral giving the residue $a_{-1}$ of the Laurent series $\sum_{p=-\infty}^N a_p z^p$.

For any $n\le 1$, the vector subspace generated by the pseudo-differential operators $D=f_n(z)\partial_z^n+f_{n-1}(z)
\partial_z^{n-1}+\ldots$ of degree $\le n$   is a Lie subalgebra of $\PsiD$ that we shall denote by $\PsiD_{\le n}$. 
We shall sometimes write
$D=O(\partial_z^n)$ for a pseudodifferential operator of degree $\le n$. Also, letting
$OD=\PsiD_{\ge 0}=\{ \sum_{k=0}^n f_k(z)\partial_z^k,\quad n\ge 0\}$
(differential operators) and $\volt=\PsiD_{\le -1}$ (called: {\it Volterra algebra}),
we shall denote by $(D_+,D_-)$ the decomposition of $D\in\PsiD$ along the
direct sum $OD\oplus\volt$, and call $D_+$ the {\em differential part} of $D$.

We shall also need to introduce 
the following 'extended' algebra of formal pseudodifferential symbols.

\begin{Definition}[algebra of extended pseudodifferential symbols]
{\it

Let $\DPsiD$ be the extended pseudo-differential  algebra generated as an associative algebra
 by $\xi, \xi^{-1}$ and $\partial_{\xi}^{\half},
\partial_{\xi}^{-\half}$.
}
\end{Definition}

Let $D\in\DPsiD$. As in the case of the usual algebra of pseudodifferential symbols, we shall write $D=O(\partial_z^{\kappa})$ $(\kappa\in\half\Z)$
for an extended pseudodifferential symbol with degree $\le \kappa$, and denote by $\DPsiD_{\le \kappa}$  the 
Lie subalgebra span$(f_j(\xi)\partial_{\xi}^j\ ;\ j=\kappa,\kappa-\half,\kappa-1,\ldots)$
 if $\kappa\le 1$.

The Lie algebra $\DPsiD$ contains two interesting subalgebras for our purposes:

\begin{itemize}

\item[(i)] span$(f_1(\xi)\partial_{\xi},f_0(\xi); \ f_1,f_0\in C^{\infty}(S^1))$ which is isomorphic
to $\Vect(S^1)\ltimes {\cal F}_0$;
\item[(ii)] $\DPsiD_{\le 1}:=$span$(f_{\kappa}(\xi)\partial_{\xi}^{\kappa};\ 
  \kappa=1,\half,0,-\half,\ldots, f_{\kappa}\in C^{\infty}(S^1))$, which is also the Lie algebra 
generated by span$(f_1(\xi)\partial_{\xi},
f_{\half}(\xi)\partial_{\xi}^{1/2}, f_0(\xi); \  f_1,f_{\half},f_0\in C^{\infty}(S^1))$.
\end{itemize}

As mentioned in the Introduction, the Schr\"odinger-Virasoro Lie algebra $\sv$ is isomorphic to a subquotient of 
$\DPsiD$:

\begin{Lemma}[$\sv$ as a subquotient of $\DPsiD$](see \cite{HenUnt06})

{\it Let $p$ be the projection of $\DPsiD_{\le 1}$ onto $\DPsiD_{\le 1}/\DPsiD_{\le -\half}$, and
$j$ be  the linear morphism from $\sv$ to $\DPsiD_{\le 1}$ defined by 
\BEQ {\cal L}_f\longrightarrow -\frac{\II}{2{\cal M}} f(-2\II{\cal M}\xi)\partial_{\xi},
\quad {\cal Y}_g\longrightarrow -g(-2\II{\cal M}\xi)\partial_{\xi}^{\half},
\quad {\cal M}_h\longrightarrow \II{\cal M} h(-2\II{\cal M}\xi).\EEQ
Then the composed morphism $p\circ j:\sv\to \DPsiD_{\le 1}/\DPsiD_{\le -\half}$ is  a Lie algebra isomorphism.
}

\label{lem:sv-DPsiD}
\end{Lemma}

{\bf Proof.} Straightforward computation. (Formulas look simpler with the
normalization $-2\II{\cal M}=1$.) \hfill \eop

\vskip 1 cm

It turns out that a certain non-local transformation gives an isomorphism between $\DPsiD$ and $\PsiD$. For the sake of the
reader, we shall in the sequel add the name of the variable as an index when speaking of
  algebras of (extended or not)
 pseudodifferential symbols.

\begin{Definition}[non-local transformation $\Theta$]

{\it Let $\Theta: \DPsiDxi \to \PsiDr$ be the associative algebra isomorphism defined by
\BEA
&& \partial_{\xi}^{\half} \to \partial_r, \quad \partial_{\xi}^{-\half} \to \partial_r^{-1} \nonumber\\
&& \xi \to \half r\partial_r^{-1}, \quad \xi^{-1}\to 2\partial_r r^{-1} 
\EEA
} \label{def:Theta}
\end{Definition}

The inverse morphism $\Theta^{-1}:\partial_r\to \partial_{\xi}^{\half}$, $r\to 2\xi\partial_{\xi}^{\half}$ is easily
seen to  be an algebra isomorphism because the defining relation $[\partial_r,r]=1$ is preserved by $\Theta^{-1}$.
It may be seen formally as the   
 integral transformation
$ \psi(r) \to \tilde{\psi}(\xi):= \int_{-\infty}^{+\infty} \frac{e^{- r^2/4\xi}}{\sqrt{\xi}} \psi(r)\ dr$
(one verifies straightforwardly for instance that $r\partial_r \psi$ goes to $2\xi\partial_{\xi} \tilde{\psi}$
and that $\partial_r^2 \psi$ goes to $\partial_{\xi} \tilde{\psi}$). In other words, assuming $\psi\in L^1(\R)$,
one has $\tilde{\psi}(\xi)=(P_{\xi}\psi)(0)$ $(\xi\ge 0)$ where $(P_{\xi},\xi\ge 0)$ is the usual
heat semi-group. Of course, this does not make sense at all for $\xi<0$.

{\bf Remark.} Denote by ${\cal E}_r=[r\partial_r,.]$ the Euler operator.  Let $\PsiD_{(0)}$, resp.
$\PsiD_{(1)}$ be the vector spaces generated by the operators $D\in\PsiD$ such that  ${\cal E}_r (D)=nD$ where $n$ is even, resp. odd. Then
$\PsiD_{(0)}$ is an (associative) subalgebra of $\PsiD$, and one has
$$[\PsiD_{(0)},\PsiD_{(0)}]=\PsiD_{(0)},\quad [\PsiD_{(0)},\PsiD_{(1)}]=\PsiD_{(1)},\quad [\PsiD_{(1)},\PsiD_{(1)}]=\PsiD_{(0)}.$$
Now, the inverse image of $D\in\PsiD_r$ by $\Theta^{-1}$ belongs to $\PsiD_{\xi}\subset\DPsiD_{\xi}$
 if and only if $D\in(\PsiD_r)_{(0)}$.

\begin{Lemma}[pull-back of Adler's trace]
{\it
  The pull-back by $\Theta$ of Adler's trace on $\PsiDr$  yields a trace on $\DPsiD$ defined by
\BEQ
\Tr_{\DPsiD_{\xi}}(a(\xi)\partial_{\xi}^q)\ :=\  Tr_{\PsiD_r} \left( \Theta(a(\xi)\partial_{\xi}^q) \right)
 =2 \del_{q,-1}\ . \frac{1}{2\II \pi}
\oint a(\xi) d\xi. \label{ThetaTr}
\EEQ
}
\end{Lemma}

{\bf Proof.}

Note first  that  the Lie bracket of $\PsiD_r$, resp.
$\DPsiD_{\xi}$ is graded with respect to the adjoint action of the Euler operator ${\cal E}_r:=[r\partial_r,.]$,
 resp. ${\cal E}_{\xi}:=[\xi \partial_{\xi},.]$, and that $\Theta\circ {\cal E}_{\xi}=\half {\cal E}_r\circ \Theta$.
 Now $\Tr_{\PsiD_r} D=0$ if $D\in\PsiD_r$ is not
homogeneous of degree 0 with respect to ${\cal E}_r$, hence the same is true for $\Tr_{D\PsiD_{\xi}}$. Consider
$D:=\xi^j \partial_{\xi}^{j}=\Theta^{-1}((\half r\partial_r^{-1})^j \partial_r^{2j})$: then $ \Tr_{\DPsiD_{\xi}}
 (D)=0$ if
$j\ge 0$ because (as one checks easily by an explicit computation) $\Theta(D)\in OD$;  and $\Tr_{\DPsiD_{\xi}}
(D)=0$ if
$j\le -2$ because $\Theta(D)=O(\partial_r^{-2})$.

\hfill \eop

In order to obtain time-dependent equations, one needs to add an extra dependence on a formal parameter $t$
of all the algebras we introduce. One obtains in this way loop algebras, whose formal definition is as follows:

\begin{Definition}[loop algebras]
{\it 
Let $\g$ be a Lie algebra. Then the {\em loop algebra} over $\g$ is the Lie algebra
\BEQ \loo_t \g:=\g[t,t^{-1}]].\EEQ
}
\end{Definition}

Elements of $\loo_t \g$ may also be considered as Laurent series
$\sum_{n=-\infty}^N t^n X_n$ $(X_n\in\g)$,
 or simply as functions  $t\to X(t)$, where $X(t)\in\g$.

The transformation $\Theta$ yields immediately (by lacing with respect to the time-variable $t$)
 an
algebra isomorphism
\BEQ \loo_t \Theta: \LDPsiDxi \to \loo_t (\PsiD_r), \quad D \to (t\to   \Theta(D(t))).\EEQ


\section{Time-shift transformation and symmetries of the free Schr\"odinger equation}


In order to define extended symmetries of the Schr\"odinger equation, one must first introduce
 the following time-shift
transformation.

\begin{Definition}[time-shift transformation ${\cal T}_t$]

{\it Let ${\cal T}_t: \DPsiDxi\to \LDPsiDxi$ be the linear transformation defined by
\BEQ {\cal T}_t \left( f(\xi)\partial_{\xi}^{\kappa} \right) = ({\cal T}_t f(\xi)) \partial_{\xi}^{\kappa} \EEQ
where:
\BEQ {\cal T}_t P(\xi)=P(\frac{\II}{2{\cal M}}t+\xi)\EEQ
for {\em polynomials $P$}, and
\BEQ {\cal T}_t \xi^{-k}=(\frac{\II}{2{\cal M}}t+\xi)^{-k}:=\left( \frac{\II}{2{\cal M}}t\right)^{-k}
 \sum_{j=0}^{\infty} (-1)^j \frac{k(k+1)\ldots (k+j-1)}{j!}
(-2\II{\cal M}\xi/t)^j.\EEQ

In other words, for any Laurent series $f\in\C[\xi,\xi^{-1}]]$,
 \BEQ {\cal T}_t f(\xi)=\sum_{k=0}^{\infty} f^{(k)}(\frac{\II}{2{\cal M}} t) \frac{\xi^k}{k!}. \EEQ

Then ${\cal T}_t$ is an injective Lie algebra homomorphism, with left inverse ${\cal S}_t$ given by
\BEQ {\cal S}_t (g(t,\xi))=\frac{1}{2\II\pi} \oint g(-2\II{\cal M}\xi,t) \frac{dt}{t}. \EEQ
}

\end{Definition}

{\bf Proof.} Straightforward.

\hfill \eop

\vskip 1 cm

Now comes an essential remark (see Introduction) which we shall first explain in an informal way.
The free Schr\"odinger equation $\Del_0\psi:=(-2\II{\cal M}\partial_t-\partial_r^2)\psi=0$ 
reads in the 'coordinates' $(t,\xi)$
\BEQ (-2\II{\cal M}\partial_t-\partial_{\xi})\tilde{\psi}(t,\xi)=0. \label{eq:Scheqxi} \EEQ
In the complex coordinates $z=t-2\II{\cal M}\xi$, $\bar{z}=t+2\II{\cal M}\xi$,
 one  simply gets (up to a constant) the $\bar{\partial}$-operator, whose
algebra of Lie symmetries is span$(f(t-2\II{\cal M}\xi)\partial_{\xi},g(t-2\II{\cal M}\xi)\partial_t)$
 for arbitrary functions $f,g$.
An  easy but crucial consequence of these considerations is the following:

\begin{Definition}[${\cal X}_f^{(i)}$-generators and $\Theta_t$-homomorphism]

{\it Let, for $f\in \C[\xi,\xi^{-1}]]$ and $j\in\half \Z$,
\BEQ {\cal X}_f^{(j)}=\Theta_t(-f(-2\II{\cal M}\xi)\partial_{\xi}^j) \in \loo_t(\PsiD_r) \EEQ
where $\Theta_t$ is the Lie algebra homomorphism obtained by
 composition of the time-laced non-local transformation $\loo_t(\Theta)$ and the time-shift ${\cal T}_t$,
\BEQ \Theta_t:=\loo_t(\Theta)\circ {\cal T}_t. \EEQ
}  \label{def:Theta-t}

\end{Definition}

In other words, 
\BEQ {\cal X}_f^{(j)}=\loo_t(\Theta)(-f(t-2\II{\cal M}\xi)\partial_{\xi}^j)=-f(t-\II{\cal M}
r\partial_r^{-1})\partial_r^{2j} \EEQ
 (at least
if $f$ is a polynomial). The homomorphism $\Theta_t$ will play a key role in the sequel.

\medskip

\begin{Lemma}[invariance of the Schr\"odinger equation] \label{lem:invariance}

\begin{itemize}
\item[(i)]

The free Schr\"odinger equation $\Del_0\psi(t,r)=0$ is invariant under
the Lie algebra of transformations generated by ${\cal X}^{(i)}_f$, $i\in\half\Z$ .

\item[(ii)]

Denote by $\dot{f},\ddot{f},\frac{d^3 f}{dt^3}$ the 
 time-derivatives of $f$ of order 1, 2, 3, then
\BEA
&& {\cal X}^{(1)}_f=-f(t)\partial_r^2+\II{\cal M} \dot{f}(t) r\partial_r+\half {\cal M}^2 \ddot{f}(t)  r^2
- \left( \half {\cal M}^2 \ddot{f}(t) r+\frac{\II}{6} {\cal M}^3 \frac{d^3 f}{dt^3} r^3 \right) \partial_r^{-1} +O(\partial_r^{-2});
\nonumber \label{eq:Xf} \\
\EEA
\BEA
&& {\cal X}^{(1/2)}_g=-g(t)\partial_r+\II{\cal M} \dot{g}(t) r+\frac{{\cal M}^2}{2} \ddot{g}(t) r^2 \partial_r^{-1}
+O(\partial_r^{-2});  \label{eq:Xg} \\
&& {\cal X}^{(0)}_h=- h(t)+O(\partial_r^{-1}). \label{eq:Xh}
\EEA

In particular, denoting by $D_+$ the differential part
 of a pseudo-differential operator $D$, i.e. its projection 
onto OD, the operators $({\cal X}^{(1/2)}_g)_+,({\cal X}^{(0)}_h)_+$ coincide (see Definition \ref{def:pi})
with $d\pi_0({\cal Y}_g)$, resp. $d\pi_0({\cal M}_h)$, while
\BEQ 2\II{\cal M} d\pi_0({\cal L}_f)=({\cal X}^{(1)}_f)_+ -f(t)\left( 2\II{\cal M}\partial_t-\partial_r^2\right).\EEQ

\end{itemize}

\end{Lemma}

{\bf Proof.} 
\begin{itemize}
\item[(i)] One has
\BEQ (\loo_t(\Theta))^{-1}({\cal X}_f^{(j)})={\cal T}_t \left(\xi\to f(-2\II{\cal M}\xi).\partial_{\xi}^j \right)=
\sum_{k=0}^{\infty} f^{(k)}(\frac{\II}{2{\cal M}}t) \frac{(-2\II{\cal M}\xi)^k}{k!} \ .\ \partial_{\xi}^j,\EEQ
which is easily seen by a straightforward computation to commute with the Schr\"odinger operator
$(\loo_t(\Theta))^{-1}(-2\II{\cal M}\partial_t-\partial_r^2)=-2\II{\cal M}\partial_t-\partial_{\xi}$, hence preserves the
free Schr\"odinger equation. Note that, when $f$ is a polynomial, $(\loo_t(\Theta))^{-1}({\cal X}_f^{(j)})=
-f(t-2\II{\cal M}\xi)\partial_{\xi}^j$ obviously commutes with $-2\II{\cal M}\partial_t-\partial_{\xi}$, see
eq. (\ref{eq:Scheqxi}) and following lines.
\item[(ii)] Straightforward computations. 
\end{itemize}\hfill \eop

\medskip

In other words (up to constant multiplicative factors), the projection $({\cal X}^{(k)}_f)_+$
of ${\cal X}^{(k)}_f$, $k=1,\half,0$ onto $OD$ forms a Lie algebra which coincides with   the 
realization $d\pi_0$ of the Schr\"odinger-Virasoro algebra , 
{\it apart} from the fact that $-2\II{\cal M}\partial_t$ is substituted by
 $\partial_r^2$ in the formula for ${\cal X}_f^{(1)}$. This discrepancy is not too alarming since $-2\II{\cal M}
\partial_t\equiv\partial_r^2$
 on the kernel of the free Schr\"odinger operator. As we shall see below, one may alter the ${\cal X}_f^{(1)}$ in order
 to make them 'begin with'  $-f(t)\partial_t$ as expected, but then the ${\cal X}^{(1)}_f$ appear to have a specific algebraic status.


\section{From central cocycles of $(\PsiD_r)_{\le 1}$ to the Kac-Moody type  algebra $\g$}


The above symmetry generators of the free Schr\"odinger equation, ${\cal X}_f^{(i)}$, $i\ge 1$ may be seen as
elements of ${\cal L}_t(\PsiD_r)$. The original idea (following the scheme for Hill operators recalled in the
Introduction) was to try and embed the space of Schr\"odinger operators ${\cal S}^{lin}$ into the dual of ${\cal L}_t
(\PsiD_r)$ and realize the action $d\tilde{\sigma}_{\mu}$ of Proposition \ref{prop:sigma} as part of the coadjoint representation
of an appropriate central extension of ${\cal L}_t(\PsiD_r)$. 

Unfortunately this scheme is a little too simple: it allows to retrieve only the action of the $Y$- and $M$-generators,
as could have been expected from the remarks at the end of section 3. It turns out that the ${\cal X}_f^{(i)}$, $i\le\half$
may be seen as elements of $\LPsiDr$, while the realization $d\pi_0({\cal L}_f)$ (see Definition
\ref{def:pi}) of the  generators in $\Vect(S^1)\subset\sv$
 involve {\it outer derivations} $-f(t)\partial_t$, $f\in C^{\infty}(S^1)$
 of this looped algebra. Then the above scheme works correctly, provided one
chooses the right central extension of $\LPsiDr$. As explained below, there are many possible families
of central extensions, and the correct one is obtained by 'looping' a cocycle $c_3\in H^2((\PsiD_r)_{\le 1},\R)$
which does {\it not} extend to the whole Lie algebra $\PsiD_r$.

In this section and the following ones, we shall formally assume the coordinate $r=e^{\II\theta}$ to be
on the circle $S^1$. If $f(r)=\sum_{k\in\Z} f_k r^k$, the Cauchy integral $\frac{1}{2\II\pi}
\oint_{S^1} f(r)dr$ selects the residue $f_{-1}$. Alternatively, we shall sometimes use the
angle coordinate $\theta$ in the next paragraph, so $f(r)=\sum_{k\in\Z} f_k e^{\II k\theta}$ may
be seen as a $2\pi$-periodic function.


\subsection{Central cocycles of $(\PsiD_r)_{\le 1}$}


We shall (almost) determine $ H^2(\PsiD_{\le 1},\R)$, using its natural semi-direct product structure
$\PsiD_{\le 1}=\Vect(S^1)\ltimes\PsiD_{\le 0}$. We choose to work with periodic functions
$f=f(\theta)$ in the paragraph.

One has (by using  the Hochschild-Serre spectral sequence, see for instance \cite{Fuk}):

\BEQ H^2(\PsiD_{\le 1},\R)=H^2(\Vect(S^1),\R)\oplus  H^1(\Vect(S^1),H^1(\PsiD_{\le 0},\R))
\oplus  Inv_{\Vect(S^1)}H^2 (\PsiD_{\le 0},\R). \EEQ

The one-dimensional space  $H^2(\Vect(S^1),\R)$ is generated by
 the Virasoro cocycle, which we shall denote by $c_0$.

For the second piece, elementary computations give
$[\PsiD_{\le 0}, \PsiD_{\le 0}]=\PsiD_{\le -2}$.
So $H_1(\PsiD_{\le 0},\R)$ is isomorphic to $\PsiD_{\le 0}/\PsiD_{\le -2}$, i.e. to the space of
 symbols of type $f_0+f_{-1}\partial^{-1}$. In terms of density modules,
 one has $H_1(\PsiD_{\le 0})={\cal F}_{0} \oplus {\cal F}_{-1}$.
So $  H^1(\PsiD_{\le 0},\R)=({\cal F}_{0}\oplus {\cal F}_{-1})^*=
   {\cal F}_{-1} \oplus {\cal F}_{0}$ by the standard duality 
(see section 1) ${\cal F}_{\mu}^*\simeq{\cal F}_{-1-\mu}$,
 and $H^1(\Vect(S^1), H^1(\PsiD_{\le 0},\R))=H^1(\Vect(S^1), {\cal F}_{-1} \oplus {\cal F}_{0})=
H^1(\Vect(S^1), {\cal F}_{-1})\oplus H^1(\Vect(S^1), {\cal F}_{0})$.
From the results of Fuks \cite{Fuk}, one knows that $ H^1(\Vect(S^1), {\cal F}_{1})$ is one-dimensional,
generated by
$f\partial_{\theta} \longrightarrow f ''d\theta$, and $ H^1(\Vect(S^1), {\cal F}_{0})$ is two-dimensional, generated by
$f\partial_{\theta} \longrightarrow f $ and $f\partial_{\theta} \longrightarrow f '$.
So we have proved that $H^1(\Vect(S^1), H^1(\PsiD_{\le 0},\R))\hookrightarrow
H^2(\PsiD_{\le 1},\R)$ is three-dimensional, with generators
$c_1$, $c_2$ and $c_3$ as follows:
\BEA
&& c_1(g\partial_{\theta}, \sum_{k=-\infty}^1    
f_k\partial_{\theta}^k)=\frac{1}{2\pi} \int g'' f_0 \ d\theta \\
&& c_2(g\partial_{\theta}, \sum_{k=-\infty}^1    
f_k\partial_{\theta}^k)=\frac{1}{2\pi} \int g f_{-1} \ d\theta \\
&& c_3(g\partial_{\theta}, \sum_{k=-\infty}^1    
f_k\partial_{\theta}^k)=\frac{1}{2\pi} \int g' f_{-1} \ d\theta 
\EEA

Let us finally  consider the third piece $Inv_{\Vect(S^1)} H^2(\PsiD_{\le 0},\R)$.We shall once more make use of a decomposition into a semi-direct product: setting $\volt=\PsiD_{\le -1}$, one has
$\PsiD_{\le 0}$=
${\cal F}_{0}\ltimes \volt$, where ${\cal F}_{0}$ is considered as an abelian Lie algebra, acting
 non-trivially on $\volt$. We do not know how to compute the cohomology of $\volt$,
 because of its "pronilpotent" structure, but we shall make the following:

{\bf Conjecture}:
\BEQ Inv_{{\cal F}_{0}}H^2 (\volt,\R)=0. \label{conj} \EEQ

\medskip

We shall now work out the computations modulo this conjecture.

One first gets $ H^2(\PsiD_{\le 0},\R)=H^2({\cal F}_{0},\R)\oplus H^1({\cal F}_{0}, H^1(\volt,\R))$.
Then $Inv_{\Vect(S^1)} H^2(\PsiD_{\le 0},\R)=Inv_{\Vect(S^1)} H^2({\cal F}_{0},\R)\oplus Inv_{\Vect(S^1)}
 H^1({\cal F}_{0}, H^1(\volt,\R))$. Since
${\cal F}_{0}$ is  abelian, one has $H^2({\cal F}_{0},\R)=\Lambda^2 ({\cal F}_{0}^\ast )$, and
$Inv_{\Vect(S^1)} (\Lambda^2 ({\cal F}_{0}^\ast ))\hookrightarrow H^2(\PsiD_{\le 1},\R)$
 is one-dimensional, generated by the well-known
 cocycle
\BEQ c_4(f, g)=\frac{1}{2\pi} \int(g' f-f'g) \ d\theta. \EEQ

A direct computation then shows that $[\volt, \volt]=\PsiD_{\le -3}$
, so $H_1(\volt,\R)={\cal F}_{-1}\oplus{\cal F}_{-2}$ and $H^1(\volt,\R)={\cal F}_{0}\oplus{\cal F}_{1}$ 
as $\Vect(S^1)$-module. Then $H^1({\cal F}_{0}, H^1(\volt,\R))$ is easily determined by direct 
 computation, as well as $Inv_{\Vect(S^1)} H^1({\cal F}_{0}, H^1(\volt,\R))\hookrightarrow
H^2(\PsiD_{\le 1},\R)$; the latter is
 one-dimensional, generated by the following cocycle:

\BEQ c_5(g, \sum_{k=-\infty}^1    
f_k\partial_{\theta}^k)=\frac{1}{2\pi} \int g f_{-1}\ d\theta, \EEQ

Let us summarize our results in the following:

\begin{Proposition}

{\it Assuming conjecture (\ref{conj}) holds true, the space $H^2(\PsiD_{\le 1},\R)$ is six-dimensional, generated by the cocycles $c_i$,  $i=0,\ldots,5$, defined above.}
\end{Proposition}

\vskip 0.5 cm

{\bf Remarks:}

\begin{enumerate}
\item
If conjecture (\ref{conj}) turned out to be false,
 it could only add some supplementary generators;
in any case, we have proved that $H^2(\PsiD_{\le 1},\R)$ is at least six-dimensional.

\item 
The natural inclusion $i:\PsiD_{\le 1}\longrightarrow\PsiD$ induces 
$i^\ast:H^2(\PsiD,\R)\longrightarrow H^2(\PsiD_{\le 1},\R)$; one may then
 determine the image by $i^\ast$ of the two generators of $H^2(\PsiD,\R)$
 determined by B. Khesin and O. Kravchenko \cite{KhKr91}. Set
$c_{KK_1}(D_1, D_2)=\Tr [\log \theta, D_1] D_2$ and 
$c_{KK_2}(D_1, D_2)=\Tr  [\log \partial_{\theta}, D_1] D_2$. Then $i^\ast c_{KK_1}=c_2$
and $i^\ast c_{KK_2}=c_0+c_1+c_4$.
\end{enumerate}

\bigskip

The right cocycle for our purposes turns out to be $c_3$: coming back to the radial
coordinate  $r$,  one gets
a  centrally extended Lie algebra of  pseudodifferential symbols
$\wit{\PsiD}_{\le 1}$ as follows.

\begin{Definition}

{\it

Let $\wit{\PsiD}_{\le 1}$ be the central extension of $\PsiD_{\le 1}$ associated with the cocycle $c c_3$ $(c\in\R)$, where
$c_3:\Lambda^2 \PsiD_{\le 1}\to \C$ verifies
\BEQ c_3(f\partial_r,g\partial_r^{-1})=c_3(f\partial_r^{-1},g\partial_r)=\frac{1}{2\II\pi} \oint f'g\ dr \EEQ
(all other relations being trivial).
}
\label{def:PsiDtilde}
\end{Definition}


\subsection{Introducing the Kac-Moody type Lie algebra $\g$}


Let us introduce now the looped algebra $\LPsiDrtilde$ in order to allow for time-dependence.
An element of $\LPsiDrtilde$ is a  pair
$(D(t),\alpha(t))$ where $\alpha\in \C[t,t^{-1}]]$ and $D(t)\in \LPsiDr$. By a slight
abuse of notation, we shall write $c_3(D_1,D_2)$ ($D_1,D_2\in\LPsiDr$) for the function
$t\to c_3(D_1(t),D_2(t))$, so   now   $c_3$ has to be seen as a
function-valued central cocycle
of $\LPsiDr$. In other words, we consider the looped version of the exact sequence
\BEQ 0 \ \longrightarrow\ \R\ \longrightarrow \ \widetilde{(\PsiDr)_{\le 1}} \ \longrightarrow (\PsiDr)_{\le 1}
\ \longrightarrow \ 0,\EEQ namely,
\BEQ 0 \ \longrightarrow\ \R[t,t^{-1}]]\ \longrightarrow \ \LPsiDrtilde  \ \longrightarrow \LPsiDr
\ \longrightarrow \ 0.\EEQ

As mentioned in the Introduction, $\LPsiDrtilde$ is naturally equipped with the Kac-Moody
cocycle $\LPsiDrtilde\times\LPsiDrtilde\to\C, \left( (D_1(t),\lambda_1(t)),(D_2(t),\lambda_2(t))
\right)\to \Tr D_1(t) \dot{D}_2(t)$. However this further central extension is irrelevant here.
On the other hand, we shall need to incorporate into our scheme {\em time derivations} $f(t)\partial_t$ (which are outer Lie derivations of $\LPsiDrtilde$, as is the case of any looped algebra),
obtaining thus a Lie algebra $\g$ which is the main object of this article.

\begin{Definition}[Kac-Moody type Lie algebra $\g$] \label{def:g}

Let $\g\simeq \Vect(S^1)_t\ltimes \LPsiDrtilde$ be the Kac-Moody type Lie algebra obtained from
$\LPsiDrtilde$ by including the outer Lie derivations
\BEQ f(t)\partial_t\ . \ \left(D(t),\alpha(t)\right)\ =\  \left(f(t)\dot{D}(t),f(t)\dot{\alpha}(t)\right).\EEQ

\end{Definition}


\section{Construction of the embedding $I$ of $(\DPsiDxi)_{\le 1}$ into  $\g$}


This section, as explained in the introduction to section 4,
 is devoted to the construction of an explicit embedding, denoted by $I$,
of the abstract algebra of extended pseudodifferential symbols $(\DPsiDxi)_{\le 1}$ into
 $\g$.  Loosely
speaking, the image $I((\DPsiDxi)_{\le 1})$ is made up of the ${\cal X}_f^{(j)}$, $j\le \half$ and the
${\cal X}_f^{(1)}$ {\em with  $\partial_r^2$ substituted by $-2\II{\cal M}\partial_t$} (see end of section 3). More precisely, $I\big|_{(\DPsiDxi)_{\le \half}}$ maps an operator $D$ into its
image by $\Theta_t$  (see Definition \ref{def:Theta-t}),
 namely, $\Theta_t(D)$, viewed as an element of the centrally extended Lie
algebra $\LPsiDrtilde$. On the other hand, the operator of degree one $-f(-2\II{\cal M}\xi)\partial_{\xi}$ will not be mapped to $\Theta_t(-f(-2\II{\cal M}\xi)\partial_{\xi})={\cal X}_f^{(1)}$ (which
is of degree $2$ in $\partial_r$), but to some element in the product $\g=\Vect(S^1)_t\ltimes
\LPsiDrtilde$ with both components non-zero, as described in the following

\medskip

\begin{Theorem}[homomorphism $I$]

{\it
Let $I: (\DPsiDxi)_{\le 1} \simeq \Vect(S^1)_{\xi}\ltimes (\DPsiDxi)_{\le \half} \hookrightarrow
\g=\Vect(S^1)_t \ltimes \LPsiDrtilde$ be the mapping defined by
\BEQ I\left((0,D)\right)=\left(0,\Theta_t(D)\right);  \label{eq:th5.1-1} \EEQ
\BEQ I\left( (-\frac{\II}{2{\cal M}}f(-2\II{\cal M}\xi)\partial_{\xi},0)\right)=\left(
-f(t)\partial_t,
\frac{\II}{2{\cal M}} ({\cal X}_f^{(1)})_{\le 1} \right)  \label{eq:th5.1-2} \EEQ
where 
\BEQ ({\cal X}_f^{(1)})_{\le 1}=\left(\Theta_t(-f(-2\II{\cal M}\xi)\partial_{\xi})\right)_{\le 1} =
\II{\cal M}\dot{f}(t)r\partial_r+\half {\cal M}^2 \ddot{f}(t)r^2-
 \left( \half {\cal M}^2 \ddot{f}(t) r+\frac{\II}{6} {\cal M}^3 \frac{d^3 f}{dt^3} r^3 \right) \partial_r^{-1}+\ldots
\label{eq:53} \EEQ
 (see Lemma \ref{lem:invariance}) 
is ${\cal X}_f^{(1)}$ shunted of its term of order $\partial_r^2$, i.e. the projection of ${\cal X}_f^{(1)}$
onto ${\cal L}_t((\PsiD_r)_{\le 1})$.

Then $I$ is a Lie algebra homomorphism.
}
\label{th:I}
\end{Theorem}

{\bf Proof.}

First of all, the cocycle $c_3$ (see Definition \ref{def:PsiDtilde})
vanishes on the product of two operators of the form
$\frac{\II}{2{\cal M}} ({\cal X}_f^{(1)})_{\le 1}+\Theta_t(D)$ belonging to the image
of $(\DPsiDxi)_{\le 1}$ by $I$, see eq. (\ref{eq:th5.1-1}) and (\ref{eq:th5.1-2}), because these
involve only non-negative powers of $r$. Hence $I$
may be seen as a map ($\bar{I}$, say) with values in $\Vect(S^1)_t\ltimes \LPsiDr$ (discarding the central extension).
 Now the Lie bracket $\left[ (-f_1(t)\partial_t,W_1),(-f_2(t)\partial_t,W_2)\right]$ in
$\Vect(S^1)_t\ltimes \LPsiDr$ coincides with the usual Lie bracket of the Lie algebra
$\PsiD_{t,r}$ of  pseudo-differential symbols in two
variables, $t$ and $r$, hence $I((\DPsiDxi)_{\le 1})$ may be seen as sitting in $\PsiD_{t,r}$. Then 
 \BEQ \bar{I}(-\frac{\II}{2{\cal M}} f(-2\II{\cal M}\xi)\partial_{\xi})=-f(t)\partial_t+\frac{\II}{2{\cal M}}
({\cal X}_f^{(1)})_{\le 1}=\frac{\II}{2{\cal M}} \left( {\cal L}'_f+{\cal X}_f^{(1)}\right), \EEQ
where
 ${\cal L}'_f:=-f(t)(-2\II{\cal M}\partial_t-\partial_r^2)$ is an independent copy of $\Vect(S^1)$,
 by which we mean that $[{\cal L}'_f,{\cal L}'_g]={\cal L}'_{\{f,g\}}={\cal L}'_{f'g-fg'}$ and $[{\cal L}'_f,{\cal X}_f^{(i)}]=0$
 for all $i$. This is
 immediate in the 'coordinates' $(t,\xi)$ since $(\loo_t(\Theta))^{-1}({\cal L}'_f)=-f(t)(-2\II{\cal M}\partial_t-\partial_{\xi})$
 commutes with $(\loo_t(\Theta))^{-1}({\cal X}_f^{(i)})=-f(t-2\II{\cal M}\xi)\partial_{\xi}^i$ as shown
in Lemma \ref{lem:invariance}.  Hence $\bar{I}$ is a Lie algebra
homomorphism.

 \hfill \eop

As we shall see in the next two sections, the coadjoint representation of the semi-direct product $\g$ is
the key to define a Poisson structure on ${\cal S}^{lin}$ for which the action of $\SV$ is Hamiltonian.


\section{The action of $\sv$ on Schr\"odinger operators as a coadjoint action}


Here and in the sequel, an element of $\g=\Vect(S^1)_t\ltimes \LPsiDrtilde$ will be denoted by
$(w(t)\partial_t,(W(t,r),\alpha(t)))$ (see subsection 4.2) or simply by the
 triplet $(w(t)\partial_t;
W,\alpha(t))$. Since an element of $\h=\LPsiDrtilde$ writes $(W(t),\alpha(t))$ with $W(t)\in(\PsiD_r)_{\le 1}$ (for every fixed $t$),
 it is natural
(using Adler's trace)  to
represent an element of the restricted dual $\g^*$ as a triplet $(v(t)dt^2; V dt,a(t)dt)$ with
$v\in C^{\infty}(S^1)$, $V\in \loo_t ((\PsiD_r)_{\ge -2})$ and $a\in C^{\infty}(S^1)$. The coupling between $\g$ and its
dual $\g^*$ writes then
\BEQ \langle \left( v(t)dt^2; V dt,a(t)dt \right), \left( w(t)\partial_t;W,\alpha(t) \right) \rangle_{\g^* \times \g}=
\frac{1}{2\II\pi} \oint \left[ v(t)w(t)+ \Tr_{\PsiDr}(V(t)W(t)) +a(t)\alpha(t) \right]\ dt.\EEQ

This section is devoted to the proof of the main Theorem announced in the Introduction, which we may
now state precisely:

\begin{Theorem}
\label{th:th}
{\it
Let $(\widetilde{\PsiD}_r)_{\le 1}$ be the central extension of $(\PsiDr)_{\le 1}$ associated with 
the cocycle $cc_3$ (see Definition \ref{def:PsiDtilde}) with $c=2$;  $\h=\LPsiDrtilde$ the
corresponding looped algebra, and $\g=
\Vect(S^1)_t\ltimes \LPsiDrtilde$  the corresponding Kac-Moody type extension by outer
derivations  (see
Definition \ref{def:g}).  Let also ${\cal N}$ be the affine subspace 
$\Vect(S^1)_t^*\ltimes  \left\{\left( \left[V_{-2}(t,r)\partial_r^{-2}+V_0(t)\partial_r^0\right]dt,a(t)dt
\right)\right\}\subset \g^*$ (note that $V_0$ is assumed to be a function of $t$ only). Then:

\begin{itemize}  
\item[(i)] the coadjoint action $\ad^*_{\g}$, restricted to the image
$I((\DPsiDxi)_{\le 1})$, preserves ${\cal N}$, and quotients out into an action of $\sv$;
\item[(ii)] decompose $d\tilde{\sigma}_0(X)(a(t)\Del_0+V(t,r))$, $X\in\sv$ into $d\tilde{\sigma}_0^{op}(X)(a)\Del_0
+d\tilde{\sigma}_0^{pot}(X)(a,V)$ (free Schr\"odinger operator depending only on $a$, plus a potential depending on
$(a,V)$). Then it holds
\BEA  && \ad_{\g}^*({\cal L}_f).\left(v(t)dt^2; \left[V_{-2}(t,r)\partial_r^{-2}+V_0(t)
\partial_r^0\right] dt,a(t)dt\right)=  \left( \left[-\half \ddot{f}(\oint rV_{-2}dr)-(f\dot{v}+2\dot{f}v)
\right] dt^2; \right. \nonumber\\
&& \left. \qquad \qquad  \left[   d\tilde{\sigma}_0^{pot}({\cal L}_f)(a,V_{-2}) \partial_r^{-2}+
(-f\dot{V}_0-\dot{f}V_0+a\dot{f})\partial_r^0 \right] dt ,  d\tilde{\sigma}_0^{op}({\cal L}_f)(a) dt \right); \nonumber\\
\label{eq:61L}  \EEA
\BEQ \ad_{\g}^*({\cal Y}_g).\left(v(t)dt^2; \left[ V_{-2}(t,r)\partial_r^{-2}+V_0(t)
\partial_r^0 \right] dt,a(t)dt\right)=\left( -\dot{g}(\oint V_{-2}dr) dt^2; \left(d\tilde{\sigma}_0^{pot}
({\cal Y}_g)(a,V_{-2})
\right) \partial_r^{-2} dt ,0\right); \label{eq:61Y} \EEQ
\BEQ \ad_{\g}^*({\cal M}_h).\left(v(t)dt^2; \left[ V_{-2}(t,r)\partial_r^{-2}+V_0(t)
\partial_r^0 \right] dt,a(t)dt\right)=\left( 0; \left(d\tilde{\sigma}_0^{pot}({\cal M}_h)(a,V_{-2})\right) \partial_r^{-2}dt,0\right). \label{eq:61M} \EEQ

In other words (disregarding  the $\partial_r^0$-component in $\h^*$ and the $dt^2$-component
in $\Vect(S^1)^*$) the  restriction of the  coadjoint action
 of $ \ad^*_{\g}\big|_{\sv}$
 to ${\cal N}$
 coincides
with the infinitesimal action $d\tilde{\sigma}_{0}$ of $\sv$ on ${\cal S}^{lin}=\{a(t)(-2\II{\cal M}\partial_t
-\partial_r^2)+V_{-2}(t,r)\}$.
\end{itemize}
}
\end{Theorem}

{\bf Remark.} The term $a\dot{f}\partial_r^0$ in eq. (\ref{eq:61L}) shows that the subspace of $\cal N$ with
vanishing coordinate $V_0\equiv 0$ is not stable by the action of $\sv$. The $V_0$-component is actually important
since the terms proportional to $\cal M$ or ${\cal M}^2$ in the action $d\tilde{\sigma}_0(\sv)$, see Proposition
\ref{prop:sigma} (which are affine terms for the affine representation $d\sigma_0$) will be obtained in the next
section as the image by the Hamiltonian operator of functionals of $V_0$.

\medskip

{\bf Proof of the Theorem.}

\medskip

Recall $\sv\simeq \DPsiD_{\le 1}/\DPsiD_{\le -\half}$ (see Lemma \ref{lem:sv-DPsiD}).
 The first important remark is that the  coadjoint action $\ad^*_{\g}$  of $I\left((\DPsiDxi)_{\le 1}\right)$
 on   elements
$\left(v(t)dt^2; \left[V_{-2}(t,r)\partial_r^{-2}+V_0(t)\partial_r^0\right] dt ,a(t)dt \right)\in
{\cal N}$  quotients out into an action 
of the Schr\"odinger-Virasoro
group. Namely, let $-\kappa\le -\half$ and  $(w;W,\alpha(t))=(w(t)\partial_t;\sum_{j\le 1}
 W_j(t,r)\partial_r^j,\alpha(t))\in \g$, then
\BEA
&& \langle \ad^*_{I(f(-2\II{\cal M}\xi)\partial_{\xi}^{-\kappa})} \left( v(t)dt^2; \left[V_{-2}(t,r)\partial_r^{-2}
+V_0(t)\partial_r^0\right]dt,a(t)dt\right), (w(t)\partial_t;W,\alpha(t))\rangle_{\g^*\times\g} \nonumber\\
&& = -\langle \left(\left[V_{-2}(t,r)\partial_r^{-2}+V_0(t)\partial_r^0\right]dt,a(t)dt\right), \left[ f(t)\partial_r^{-2\kappa}+O(\partial_r^{-2\kappa-1}),\sum_{j\le 1}
W_j(t,r)\partial_r^j \right]_{ \h} \rangle_{\h^*\times\h} \nonumber\\
&&+\langle \left[ V_{-2}(t,r)\partial_r^{-2}+V_0(t)\partial_r^0\right] dt, w(t)\dot{f}(t)
\partial_r^{-2\kappa}+O(\partial_r^{-2\kappa-1}) \rangle \nonumber\\
&&=0
\EEA
since the Lie bracket in $\g$ produces (i) no term along the central charge (namely, the coefficient of
$\partial_r^{-1}$ is constant in $r$, see Definition \ref{def:PsiDtilde}); (ii) if $-\kappa=-\half$ only,
a term of order $-1$ coming from $-[f(t)\partial_r^{-1},W_1\partial_r]+w(t)\dot{f}(t)\partial_r^{-1}=
\left(f(t)W'_1+w(t)\dot{f}(t)\right) \partial_r^{-1}+\ldots$, whose coupling with
the potential yields $\int\int V_0(t) \left(f(t)W'_1+w(t)\dot{f}(t)\right) dt dr=0$ (total
derivative in $r$); 
(iii)  a pseudodifferential operator of degree $\le -2$
which does not couple to the potential. 

Denote by $p\circ j$ the isomorphism from $\sv$ to $\DPsiD_{\le 1}/\DPsiD_{\le -\half}$, as in Lemma
\ref{lem:sv-DPsiD}. With a slight abuse of notation, we shall write $\ad^*_{X}$ instead of $\ad^*_{p\circ j(X)}$ for
$X\in\sv$ and consider  $\ad^*\circ p\circ j$ as a 'coadjoint action' of $\sv$. 

Let us now study successively the ' coadjoint action' of the $Y$, $M$ and $L$ generators of $\sv$ on
 elements $\left(v(t)dt^2;\left[V_{-2}(t,r)\partial_r^{-2}+V_0(t)\partial_r^0\right] dt,a(t)dt\right)\in \g^*$.

Recall from the Introduction that 
 the derivative with respect to $r$, resp. $t$  is denoted by $'$, resp. by a dot, namely,
$V'(t,r):=\partial_r V(t,r)$ and $\dot{V}(t,r):=\partial_t V(t,r)$.

\bigskip

\underline{Action of the $Y$-generators}

Let $W=\sum_{j\le 1}
W_j(t,r)\partial_r^j\in\loo_t((\PsiD_r)_{\le 1})$ and $\alpha(t)\in C^{\infty}(S^1)$ as before. 
A computation gives (see equation (\ref{eq:Xg}))
\BEA
&&\langle \ad^*_{{\cal Y}_g} \left(v(t)dt^2;\left[V_{-2}(t,r)\partial_r^{-2}+V_0(t)\partial_r^0\right] dt ,a(t)dt
\right),(w(t)\partial_t;W,\alpha(t))\rangle_{\g^*\times\g}= \nonumber\\
&&-\langle  \left( \left[V_{-2}(t,r)\partial_r^{-2}+V_0(t)\right] dt,a(t)dt\right),
 \large[ -g(t)\partial_r+\II{\cal M}
\dot{g}(t)r+\frac{{\cal M}^2}{2} \ddot{g}(t)r^2\partial_r^{-1}+O(\partial_r^{-2}),   \nonumber\\
&&  W_1(t,r)\partial_r+W_0(t,r)\partial_r^0+W_{-1}(t,r)\partial_r^{-1}+O(\partial_r^{-2}) \large]_{\h} - w\left( -\dot{g}\partial_r+\frac{{\cal M}^2}{2} \frac{d^3 g}{dt^3} r^2 \partial_r^{-1} 
\right) \rangle_{\h^*\times\h} \nonumber\\
&&=-\langle  \left( V_{-2}(t,r)\partial_r^{-2}  dt,a(t) dt\right), \left( -(g
 W'_1-\dot{g}w)\partial_r,  c{\cal M}^2 \ddot{g} \ .\ \frac{1}{2\II\pi} \oint rW_1 dr 
\right) \rangle \nonumber \\
&& \qquad \qquad \qquad + \langle  V_0(t)\partial_r^0 dt, g(t)W'_{-1}+\left(\frac{{\cal M}^2}{2} \ddot{g} r^2
W_1\right)' \rangle \nonumber\\
&&= \int \int V_{-2} (gW'_1-\dot{g}w) \ dt\ dr-c{\cal M}^2\ . \frac{1}{2\II\pi} \int \int a\ddot{g} rW_1\ dt\ dr.
\EEA

The coupling of $V_0$ with $W$ vanishes, as may be seen in greater generality as follows (this
will be helpful later when looking at the action of the $L$-generators): 
 the term of order $-1$ comes from  a bracket of the type
$[A(t,r)\partial_r,B(t,r)\partial_r^{-1}]=(A'B+AB')\partial_r^{-1}+\ldots$; this
 is a total derivative in $r$, hence (since $V'_0\equiv 0$ by hypothesis)
$\langle V_0(t)\partial_r^0 dt, [A\partial_r,B\partial_r^{-1}]\rangle=0.$

Generally speaking (by definition of the duality given by Adler's trace),
 the terms in the above expression that depend on $W_i$, $i=1,0,\ldots$ give the projection of 
$\ad^*_{{\cal Y}_g}(v(t)dt^2; V dt,a(t)dt)$ on the component $\partial_r^{-i-1}$, while the
term depending on $w$ gives the projection on the $\Vect(S^1)$-component.

Hence altogether one has proved:

\BEQ
\ad^*_{{\cal Y}_g}\left ((v(t)dt^2; \left[V_{-2}(t,r)\partial_r^{-2}+V_0(t)\partial_r^0\right]dt,a(t)dt) \right)
=\left( -\dot{g} (\oint V_{-2}dr)dt^2; -\left(g(t)V'_{-2}+c{\cal M}^2 a
  \ddot{g}(t) r \right) \partial_r^{-2} dt,0\right)
\EEQ

which gives the expected result for $c=2$.

\bigskip

\underline{Action of the $M$-generators}

It may be deduced from that of the $Y$-generators since the Lie brackets of the $Y$-generators generate
all $M$-generators.

\bigskip

\underline{Action of the Virasoro part}

One computes (see equation (\ref{eq:Xf}) or (\ref{eq:53})): 

\BEA
&&\langle \ad^*_{{\cal L}_f} (v(t)dt^2;[V_{-2}(t,r)\partial_r^{-2}+V_0(t)\partial_r^0] dt ,a(t)dt),
(w(t)\partial_t;W,\alpha(t))\rangle_{\g^*\times\g}= \nonumber\\
&& -\langle \ad^*_{\Vect(S^1)} f(t)\partial_t. v(t)dt^2,w(t)\partial_t\rangle_{\Vect(S^1)^*\times
\Vect(S^1)} \nonumber\\
&&-\langle \left( \left[V_{-2}(t,r) \partial_r^{-2}+V_0(t)\partial_r^0\right]  dt,a(t)dt\right),
 -f(t)\partial_t\ .\  \left(W_1(t,r)\partial_r+W_0(t,r)+W_{-1}(t,r)\partial_r^{-1}+O(\partial_r^{-2})\right) \nonumber\\ 
&& +\frac{\II}{2{\cal M}} \left[ \II{\cal M} \dot{f}(t)r\partial_r+\frac{{\cal M}^2}{2}r^2 \ddot{f}(t)-\left(\frac{{\cal M}^2}{2} \ddot{f}(t) r+\frac{\II}{6} {\cal M}^3 \frac{d^3 f}{dt^3} r^3\right) \partial_r^{-1}+O(\partial_r^{-2}), \right. \nonumber\\
&& \left. \qquad \qquad \qquad  \qquad \qquad \qquad 
 W_1(t,r)\partial_r+W_0(t,r)+W_{-1}(t,r)\partial_r^{-1}+O(\partial_r^{-2}) \right]_{\h}  \rangle_{\h^*
\times \h} \nonumber\\
&& +\int dt  \ \Tr (V_{-2}(t,r)\partial_r^{-2}+V_0(t)). w(t). \left(-\half \ddot{f} r\partial_r+
\left(-\frac{\II}{4} {\cal M} \frac{d^3 f}{dt^3} r-\frac{{\cal M}^2}{12}  \frac{d^4 f}{dt^4} r^3
\right) \partial_r^{-1} \right) \nonumber\\
&&=-\int (f\dot{v}+2\dot{f}v)w dt\ +\ \  \langle \left(  V_{-2}(t,r)\partial_r^{-2} dt,a(t)dt\right), 
 \left(  \left(f(t) \dot{W}_1+\half \dot{f}(t)(rW'_1-W_1) \right)\partial_r,  \right. \nonumber\\
&& \left. \qquad \qquad \qquad -f\dot{\alpha}\ +\  \frac{c}{2\II\pi}\frac{\II}{2{\cal M}}  \left( -\II {\cal M} \dot{f}(t) \oint W_{-1} dr +\frac{{\cal M}^2}{2} \ddot{f}(t)  \oint W_1 dr +\frac{\II}{2} {\cal M}^3  \frac{d^3 f}{dt^3} 
\oint r^2 W_1 dr \right) \right)  \rangle_{\h^*\times\h} \nonumber \\
&&+ \langle V_0(t)\partial_r^0 dt,f(t)\dot{W}_{-1} \partial_r^{-1} \rangle_{\h^*\times\h} -\half
\int\int w\ddot{f}rV_{-2} \ dt\ dr.
\EEA

A term of the form $\langle  V_0(t)\partial_r^0 dt, [A\partial_r,B\partial_r^{-1}]\rangle$
 (which vanishes after integration
as above, see computations for the action of the $Y$-generators) has been left out. The term depending on $\alpha$
gives the projection on the $a$-coordinate.

Hence:

\BEA 
&& \ad^*_{{\cal L}_f}((v(t)dt^2; \left[V_{-2}(t,r)\partial_r^{-2}+V_0(t)\partial_r^0\right]dt,a(t)dt))
=  \quad  \left( \left[- \half \ddot{f} (\oint rV_{-2} dr)-(f\dot{v}+2\dot{f}v)\right]dt^2;
\right.\nonumber\\
&& \left.  \left[ \left(-f(t)\dot{V}_{-2}-\half \dot{f}(t) (rV'_{-2}+4V_{-2})+ca(t) \left(\frac{\II{\cal M}}{4}
 \ddot{f}(t) -\frac{{\cal M}^2}{4} r^2 
\frac{d^3 f}{dt^3}\right) \right)\partial_r^{-2} + \left( -f\dot{V}_0-\dot{f}V_0+\frac{c}{2}a\dot{f} \right) \partial_r^0\right]dt, \right. \nonumber\\
&& \left. \qquad \qquad \qquad \qquad \qquad \qquad \qquad \qquad \qquad \qquad \qquad \qquad  -(a\dot{f}+f\dot{a})dt \right)  
\EEA

which gives the expected result for $c=2$. \hfill \eop

{\bf Remark:} By modifiying as follows the relation defining the non-local transformation $\Theta$
(see Definition \ref{def:Theta})
\BEQ \partial_{\xi}^{\half}\longrightarrow \partial_r,\quad
\xi\longrightarrow \half r\partial_r^{-1}+\nu\partial_r^{-2} 
\EEQ
for an arbitrary real parameter $\nu$, one may obtain all the actions in the family
$d\tilde{\sigma}_{\mu}$, $\mu\in\R$ (as detailed but straightforward computations
show). Note that 
$$\left(\half r\partial_r^{-1}\right)^*=-\half \partial_r^{-1}r=-\half r\partial_r^{-1}+\half
\partial_r^{-2}$$
so the operators $\half r\partial_r^{-1}+\nu\partial_r^{-2}$, $\nu\in\R$ correspond to
various (and mainly harmless) symmetrizations of $\half r\partial_r^{-1}$.


\section{Connection with the Poisson formalism}


The previous results suggest by the Kirillov-Kostant-Souriau formalism that $d\sigma_0(X)$, $X\in \sv$ is
a Hamiltonian vector field, image of some function $F_X$ by the Hamiltonian operator.
 It is the purpose of this section to write down
properly the Hamiltonian operator $H$ and to spell out  for every $X\in \sv$ a  function $F_X$  such that
$H_{F_X}=X$.

Identify $\h^*$ as a subspace of ${\cal L}_t((\PsiDr)_{\ge -2})\oplus {\cal F}_{-1}$ through the
pairing given by Adler's trace as in the first lines of section 6, so that an element
of $\h^*$ writes generically $\left( \sum_{k\ge -2} V_k\partial_r^k \ .\ dt, a(t)dt\right)$. Consider 
similarly to \cite{GuiRog} the space ${\cal F}_{loc}$ of local functionals on ${\cal L}_t((\PsiDr)_{\ge -2})$, \\
${\cal F}_{loc}:=
\hat{\cal F}_{loc}/{\mathrm{span}}\left( \frac{d}{dt} \hat{\cal F}_{loc}, \frac{d}{dr} \hat{\cal F}_{loc}\right)$, 
with $\hat{\cal F}_{loc}=C^{\infty}(S^1\times S^1)\otimes \C[(\partial_t^i \partial_r^j V_k)_{k\ge -2,i,j\ge 0}]$.
An element $F$ of ${\cal F}_{loc}$ defines by integration 
 a $\C$-valued function $\int \int F(t,r)\ dt dr$ on ${\cal L}_t((\PsiDr)_{\ge -2})$.  The classical Euler-Lagrange variational formula yields the variational
derivative
\BEQ \frac{\del F}{\del V_k}=\sum_{i,j=0}^{\infty} (-1)^{i+j} \partial_t^i \partial_r^j \left( \frac{
\partial F}{\partial (\partial_t^i \partial_r^j V_k)} \right). \EEQ 
{\it Local vector fields} are then formally derivations of $\hat{\cal F}_{loc}$ commuting with $\frac{d}{dt}$ and
$\frac{d}{dr}$, so that they define linear morphisms $X:{\cal F}_{loc}\to {\cal F}_{loc}$. It is also possible
to represent $X$ more geometrically as a vector field on ${\cal L}_t((\PsiDr)_{\ge -2})$; since ${\cal L}_t((\PsiDr)_{\ge -2})$ is linear, $X$ is a mapping
$X:{\cal L}_t((\PsiDr)_{\ge -2})\to{\cal L}_t((\PsiDr)_{\ge -2})$ with some additional requirements due to locality.
 Set $X(D)=\sum_{k\ge -2} A_k(D) \partial^k$, then (as a 
derivation of $\hat{\cal F}_{loc}$) it holds $X=\sum_{k\in\Z} \sum_{i,j\ge 0} \partial_t^i \partial_r^j a_k \ .\
\partial/\partial(\partial_t^i \partial_r^j V_k)$. Now the differential $dF$ of a function $F\in {\cal F}_{loc}$
verifies by definition
$dF(X)=X(F)=\sum_k a_k \frac{\del F}{\del V_k}$. Choose $D\in{\cal L}_t((\PsiDr)_{\ge -2})$: then the differential of $F$ at $D$
should be a linear evaluation $\langle d_D F,X(D)\rangle=\int  Tr d_D F(t) X(D)(t)\ dt$, hence (using once again the
pairing given by Adler's trace) one has the following representation: $d_D F=\sum_k \partial^{-k-1}
\frac{\del F}{\del V_k}(D)\in \h$. Formally, one may simply write $dF=\sum_k \partial^{-k-1} \frac{\del F}{\del V_k}$.

Similar considerations apply to local functionals on $\Vect(S^1)^*$ or ${\cal F}_{-1}$, with the difference
that the variable $r$ is absent. We refer once again to \cite{GuiRog} for this very classical case.
Since the generic element of $\Vect(S^1)$, resp. $\Vect(S^1)^*$, is denoted by $w(t)\partial_t$,
resp. $v(t)dt^2$, the differential of a functional $F=F(v)$ will be denoted by
$dF=\frac{\del F}{\del v} \partial_t$, while a vector field writes $X(v)=A_{{\cal F}_{-2}}(v)
dt^2$. Similarly,  the differential of a functional $F=F(a)$ will be denoted by
$dF=\frac{\del F}{\del a}$, while a vector field writes $X(a)=A_{{\cal F}_{-1}}(a)
dt$. Note that (considering e.g. the case of $\Vect(S^1)^*$) such a functional may be seen as a particular case
of a 'mixed-type local functional' $\Phi(v,(V_k)_{k\ge -2})$ by setting $\Phi(v,(V_k)_{k\ge -2})=
r^{-1}F(v)$ (integrating with respect to $r$ yields $\oint r^{-1}dr=1$), but we shall not need
such mixed-type functionals. We shall  restrict to (i) local functionals on ${\cal L}_t((\PsiDr)_{\ge -2})$, (ii) local
functionals on $\Vect(S^1)^*$ and (iii) local functionals on ${\cal F}_{-1}$, which are sufficient for our purposes.

It is now possible to write down explicitly the Poisson bracket of local functionals of
the above three types on $\g^*$; we shall
 restrict to the affine subspace $\Vect(S^1)^*\ltimes\{(\left[V_{-2}(t,r)\partial_r^{-2}+V_0(t,r)
\partial_r^0\right]dt,a(t)dt)\}\subset\g^*$ (note that we allow a dependence on  $r$ of the potential $V_0$ for the
time being). Denote by
$V=(V_{-2},V_0)$ the element $V_{-2}(t,r)\partial_r^{-2}+V_0(t,r)\partial_r^0$. 
Consider first local functionals $F,G$ on ${\cal L}_t((\PsiDr)_{\ge -2})$. By the
Kirillov-Kostant-Souriau construction,
\BEA  && \left\{\int\int F \ dtdr ,\int\int G\ dt dr\right\}\left(\left( v(t)dt^2; \left[V_{-2}(t,r)\partial_r^{-2}+V_0(t,r)\partial_r^0\right]dt,a(t)dt\right)\right) \nonumber\\
&& \qquad \qquad \qquad =
\langle (\left[V_{-2}\partial_r^{-2}+V_0\right]dt,a(t)dt), [d_V F,d_V G]_{\h} \rangle_{\h^* \times\h} \nonumber\\
&& \qquad \qquad \qquad = \int \left\{ \Tr \left( (V_{-2}\partial_r^{-2}+V_0).[d_V F,d_V G]_{\LPsiDr} \right) + cc_3(d_V F,d_V G) \right\} \ dt. \nonumber\\
 \EEA 
Recall from the previous considerations that $d F=\partial_r \frac{\del F}{\del V_{-2}}+\frac{\del F}{\del
V_{-1}}+\partial_r^{-1} \frac{\del F}{\del V_0}+\ldots$ The operation of taking the trace leaves out only the
 bracket $\left[\partial_r \frac{\del F}{\del V_{-2}},\partial_r \frac{\del G}{\del V_{-2}}\right]$ which
couples to  $V_{-2}\partial_r^{-2} $, and the mixed
brackets  $\left[\partial_r \frac{\del F}{\del V_{-2}},\partial_r^{-1} \frac{\del G}{\del V_0}\right]$ and
$ \left[\partial_r \frac{\del G}{\del V_{-2}},\partial_r^{-1} \frac{\del F}{\del V_0}\right]$ which
couple to $V_0$, while the
central extension couples only the coefficients of $\partial_r$ and $\partial_r^{-1}$. All
together one obtains
\BEA && \{\int \int F \ dt dr,\int\int G\ dt dr\}\left(\left( v(t)dt^2; \left[V_{-2}(t,r)\partial_r^{-2}+V_0(t,r)\partial_r^0 \right]dt,a(t)dt\right)\right) \nonumber\\
&&=
\int \int V_{-2} \left[ \left( \frac{\del G}{\del V_{-2}}\right)' \frac{\del F}{\del V_{-2}} -
\left( \frac{\del F}{\del V_{-2}}\right)' \frac{\del G}{\del V_{-2}} \right] \ dt\ dr \nonumber\\
&& \qquad +  \int\int V_0\left[ \frac{\del G}{\del V_0} \frac{\del F}{\del V_{-2}}-\frac{\del G}{\del V_{-2}}
\frac{\del F}{\del V_0}\right]'\ dt\ dr \nonumber\\
&& \qquad +  c \int \int \left[ \left( \frac{\del F}{\del V_0}\right)' \  .\  \left( \frac{\del G}{\del V_{-2}} \right) + \left(
\frac{\del F}{\del V_{-2}}\right)' \ . \ \left( \frac{\del G}{\del V_0}\right) \ \right] a(t)\ dt \ dr. \nonumber\\ \EEA

Assume now that $F$ is a functional on $\Vect(S^1)^*$ and $G$ a functional on ${\cal L}_t((\PsiDr)_{\ge -2})$; then
\BEA  && \left\{\int F\ dt ,\int\int G\ dt dr \right\}\left(\left( v(t)dt^2; \left[V_{-2}(t,r)\partial_r^{-2}+V_0(t,r)\partial_r^0 \right]dt,a(t)dt\right)\right)  \nonumber\\
&& \qquad \qquad \qquad = 
\langle (\left[V_{-2}\partial_r^{-2}+V_0\right]dt,a(t)dt), \frac{\del F}{\del v}\partial_t. d_V G \rangle_{\h^*\times\h} \nonumber\\
&& \qquad \qquad \qquad =  \int \int \frac{\del F}{\del v} \ .\ \left( V_{-2} \frac{d}{dt} \left( \frac{\del G}{\del V_{-2}}
\right) +V_0 \frac{d}{dt} \left( \frac{\del G}{\del V_0}\right) \right)\ dt\ dr. \nonumber\\ \EEA

Similarly, if $F$ is a functional on $\Vect(S^1)^*$ and $G$ a function on ${\cal F}_{-1}$, then
\BEQ \left\{ \int F\ dt,\int G\ dt \right\} \left(\left( v(t)dt^2; \left[V_{-2}(t,r)\partial_r^{-2}+V_0(t,r)\partial_r^0 \right]dt,a(t)dt\right)\right)= \int a(t) \frac{\del F}{\del v} \frac{d}{dt} \left( \frac{\del G}{\del a}\right)\ dt.
\EEQ

Finally, if both $F$ and $G$ are functionals on $\Vect(S^1)^*$, then (as is classical)
\BEQ \left\{\int F\ dt,\int G\ dt\right\}\left(\left( v(t)dt^2; \left[V_{-2}(t,r)\partial_r^{-2}+V_0(t,r)\partial_r^0
\right]dt,a(t)dt\right)\right)=
\int v(t) \left[ \frac{\del F}{\del v} \frac{d}{dt}\left( \frac{\del G}{\del v}\right) -
\frac{\del G}{\del v} \frac{d}{dt}\left(\frac{\del F}{\del v}\right) \right] dt. \EEQ

Consider now the Hamiltonian operator $F\to H_F$. Set $H_F=A_{{\cal F}_{-2}} dt^2+\sum_{k\ge -2} A_k \partial^k+
A_{{\cal F}_{-1}}dt$, then
\BEQ dG(H_F)(v(t) dt^2;V dt ,a(t)dt)=H_F(G)(v(t) dt^2;V dt ,a(t)dt)=\{F,G\}(v(t) dt^2;V dt ,a(t) dt) \EEQ
writes $\int \int \sum_{k\ge -2}  A_k
 \frac{\del G}{\del V_k} (V)\ dt dr$ if $G$ is a functional on ${\cal L}_t((\PsiDr)_{\ge -2})$,  $
\int A_{{\cal F}_{-2}} \frac{\del G}{\del v}(v)\ dt$ if $G$
is a functional on $\Vect(S^1)^*$, and $\int A_{{\cal F}_{-1}} \frac{\del G}{\del a}(a) \ dt$ if $G$ if a functional
on ${\cal F}_{-1}$,
hence
\BEA &&  H_F\left(v(t)dt^2; \left[V_{-2}(t,r)\partial_r^{-2}+V_0(t,r)\partial_r^0 \right]dt,a(t)dt\right)=\left(
-\oint dr \left[ V_{-2} \frac{d}{dt} \left( \frac{\del F}{\del V_{-2}}\right)+V_0 \frac{d}{dt}
\left( \frac{\del F}{\del V_0}\right) \right] dt^2; \right. \nonumber\\
&& \quad \left.  \left(-
2V_{-2} \left( \frac{\del F}{\del V_{-2}}\right)'-V'_{-2} \frac{\del F}{\del V_{-2}}+ca(t)\left( \frac{\del F}{\del V_0}\right)' +V'_0 \frac{\del F}{\del V_0}  \right) \partial_r^{-2} + 
\left( ca(t)\left( \frac{\del F}{\del V_{-2}}\right)' -V'_0 \frac{\del F}{\del V_{-2}} \right)\partial_r^0,0 
\right) \nonumber\\ \EEA
if $F$ is a functional on ${\cal L}_t((\PsiDr)_{\ge -2})$, and 
\BEA &&   H_F\left(v(t)dt^2; \left[V_{-2}(t,r)\partial_r^{-2}+V_0(t,r)\partial_r^0\right]dt,a(t)dt\right) \nonumber\\
&& \qquad \qquad =\left(
\left[ -2v\frac{d}{dt} \left( \frac{\del F}{\del v}\right)-\dot{v} \frac{\del F}{\del v} \right]
dt^2;
-\frac{d}{dt}\left( V_{-2}\frac{\del F}{\del v}\right) \partial_r^{-2}-\frac{d}{dt}\left( V_0
\frac{\del F}{\del v} \right) \partial_r^0,-\frac{d}{dt}(a\frac{\del F}{\del v}) dt \right) \nonumber\\
\EEA

if $F$ is a functional on $\Vect(S^1)^*$.

Let $F$ be a functional on ${\cal L}_t((\PsiDr)_{\ge -2})$ depending only on $V_0$ and $V_{-2}$; note
 that $H_F$ preserves the affine subspace $\cal N$ if and only if
\BEQ F=F_0(V_0)+\sum_{i,j=0}^{\infty} \partial_t^i \partial_r^j V_{-2}. \sum_{k=0}^{j+1}
r^k f_{ijk}(V_0),\EEQ
where $F_0$ is any functional depending on $V_0$ and $t,r$, and $(f_{ijk})_{i,j,k}$ any set
of  functionals depending on $V_0$ and only on $t$.  In particular,
such a functional is affine in $V_2$ and its derivatives, and the coefficient of $V_{-2}$ affine in $r$.

\begin{Lemma}

The coadjoint action $\ad^*_{\g}(X)$ of $X={\cal L}_f$, resp. ${\cal Y}_g$, resp. ${\cal M}_h\in\sv$
 on $\cal N$ may
be identified with the Hamiltonian vector field $H_{F_X}$ with
\BEQ F_{{\cal L}_f}(v,(V_k)_{k\in\Z})=\int fv\  dt+\half\int \int  r\dot{f} V_{-2}\ dt\ dr+
\int \int \left( \II \frac{\cal M}{4} r
\ddot{f}-\frac{{\cal M}^2}{12} r^3 \frac{d^3}{dt^3} f\right) V_0 \ dt \ dr; \EEQ
\BEQ F_{{\cal Y}_g}((V_k)_{k\in\Z})= \int \int gV_{-2}\ dt \ dr-
\frac{{\cal M}^2}{2} \int \int  \ddot{g} r^2 V_0 \ dt \ dr; \EEQ
\BEQ F_{{\cal M}_h}((V_k)_{k\in\Z})= {\cal M}^2\int \int  r\dot{h}V_0 \ \ dt \ dr.\EEQ

Furthermore, $\{F_X,F_Y\}=F_{[X,Y]}$ if $X,Y\in\sv$, except for the Poisson brackets
\BEQ \{F_{{\cal Y}_g},F_{{\cal M}_h}\}={\cal M}^2 \int \int \dot{h}gV_0\ dt dr;\EEQ
\BEQ \{ F_{{\cal L}_f}, F_{{\cal Y}_g}\}=F_{[{\cal L}_f,{\cal Y}_g]}-\II \frac{\cal M}{4}
\int\int g\ddot{f} V_0\ dt dr.\EEQ
The additional terms on the right are functionals of the form $\int \int f V_0\ dt dr$ which
vanish on ${\cal N}$ (and whose Hamiltonian acts of course trivially on ${\cal N}$).

\end{Lemma}

{\bf Proof.} Straightforward computations. \hfill \eop


\small{
 }

\end{document}